\documentclass[
 reprint,
superscriptaddress,
showpacs,
 amsmath,
 amssymb,
 aps,
 prc
]{revtex4-1}

\usepackage{graphicx}
\usepackage{dcolumn}
\usepackage{hyperref}
\usepackage{physics}
\usepackage{bm}

\newcommand{\Ss}{${}^{1}\mathrm{S}_{0}$}
\newcommand{\Ps}{${}^{1}\mathrm{P}_{1}$}

\newcommand{\Pt}[1]{${}^{3}\mathrm{P}_{#1}$}

\newcommand{\Ra}{$^{225}$Ra}
\newcommand{\ra}{$^{226}$Ra}

\begin{document}


\title{Improved limit on the \Ra\ electric dipole moment}

\author{Michael Bishof}
\email{bishof@anl.gov}
\affiliation{Physics Division, Argonne National Laboratory, Argonne, Illinois 60439, USA}

\author{Richard H. Parker}%
\altaffiliation[Present address: ]{Department of Physics, University of California at Berkeley}
\affiliation{Department of Physics and Enrico Fermi Institute, University of Chicago, Chicago, Illinois 60637, USA }
\affiliation{Physics Division, Argonne National Laboratory, Argonne, Illinois 60439, USA}

\author{Kevin G. Bailey}

\author{John P. Greene}

\author{Roy J. Holt}
\affiliation{Physics Division, Argonne National Laboratory, Argonne, Illinois 60439, USA}

\author{Mukut R. Kalita}
\altaffiliation[Present address:  ]{TRIUMF, Vancouver, BC  V6T 2A3, Canada}
\affiliation{Physics Division, Argonne National Laboratory, Argonne, Illinois 60439, USA}
\affiliation{Department of Physics and Astronomy, University of Kentucky, Lexington, Kentucky 40506, USA}

\author{Wolfgang Korsch}
\affiliation{Department of Physics and Astronomy, University of Kentucky, Lexington, Kentucky 40506, USA}

\author{Nathan D. Lemke}%
\altaffiliation[Present address: ]{Space Dynamics Laboratory, Logan, Utah 84341, USA.}
\affiliation{Physics Division, Argonne National Laboratory, Argonne, Illinois 60439, USA}

\author{Zheng-Tian Lu}%
\altaffiliation[Present address: ]{University of Science and Technology of China}
\affiliation{Department of Physics and Enrico Fermi Institute, University of Chicago, Chicago, Illinois 60637, USA }
\affiliation{Physics Division, Argonne National Laboratory, Argonne, Illinois 60439, USA}

\author{Peter Mueller}

\author{Thomas P. O'Connor}
\affiliation{Physics Division, Argonne National Laboratory, Argonne, Illinois 60439, USA}

\author{Jaideep T. Singh}
\affiliation{National Superconducting Cyclotron Laboratory and Department of Physics and Astronomy,
Michigan State University, East Lansing, Michigan 48824, USA}

\author{Matthew R. Dietrich}%
\affiliation{Physics Division, Argonne National Laboratory, Argonne, Illinois 60439, USA}

\date{\today}

\begin{abstract}

\begin{description}
\item[Background] Octupole-deformed nuclei, such as that of \Ra, are expected to amplify observable atomic electric dipole moments (EDMs) that arise from time-reversal and parity-violating interactions in the nuclear medium.  In 2015, we reported the first ``proof-of-principle'' measurement of the \Ra\ atomic EDM.
\item[Purpose] This work reports on the first of several experimental upgrades to improve the statistical sensitivity of our \Ra\ EDM measurements by orders of magnitude and evaluates systematic effects that contribute to current and future levels of experimental sensitivity.
\item[Method] Laser-cooled and trapped \Ra\ atoms are held between two high voltage electrodes in an ultra high vacuum chamber at the center of a magnetically shielded environment.  We observe Larmor precession in a uniform magnetic field using nuclear-spin-dependent laser light scattering and look for a phase shift proportional to the applied electric field, which indicates the existence of an EDM.  The main improvement to our measurement technique is an order of magnitude increase in spin precession time, which is enabled by an improved vacuum system and a reduction in trap-induced heating.  
\item[Results]  We have measured the \Ra\ atomic EDM to be less than  $1.4\times10^{-23}$ $e$ cm ($95\%$ confidence upper limit), which is a factor of 36 improvement over our previous result.  
\item[Conclusions]  Our evaluation of systematic effects shows that this measurement is completely limited by statistical uncertainty.  Combining this measurement technique with planned experimental upgrades we project a statistical sensitivity at the $1\times10^{-28}$ $e$ cm level and a total systematic uncertainty at the $4\times10^{-29}$ $e$ cm level. 
\end{description}

\end{abstract}

\pacs{32.10.Dk, 11.30.Er, 24.80.+y, 37.10.Gh}
\maketitle


\section{\label{sec:introduction}Introduction}
A permanent electric dipole moment (EDM) in a non-degenerate system would violate both parity inversion (\textit{P}) and time reversal (\textit{T}) symmetries.  Assuming invariance under the combined charge conjugation (\textit{C}), \textit{P}, and \textit{T} transformations, an EDM also violates \textit{CP} symmetry.  Measurements of EDMs in a wide variety of systems, such as the neutron \cite{Baker06,Pendlebury15}, electron \cite{Baron14}, and nuclei \cite{Griffith09,Graner16}, place stringent limits on various sources of \textit{CP} violation.  For example, measurements of the neutron EDM  \cite{Baker06,Pendlebury15} have traditionally been used to restrict the CP-violating parameter of quantum chromodynamics to be $\Theta\leq10^{-10}$, whereas one would expect it to be of order unity \cite{Pospelov05}.

Identifying additional sources of \textit{CP} violation is essential to understanding how matter came to dominate over anti-matter in our universe.  Although the Standard Model (SM) includes \textit{CP} violation observed in $K$ meson \cite{Christenson64} and $B$ meson \cite{Aubert01,Abe01} decay, these mechanisms are unable to reproduce the observed matter-antimatter asymmetry \cite{Canetti12}. \textit{CP} violation included in the SM predicts EDMs that are many orders of magnitude below current experimental limits such that observation of a non-zero EDM in the foreseeable future would be a signature of new physics.

The EDM of a diamagnetic atom arises primarily from \textit{CP} violating interactions in the nuclear medium.  Although atomic electrons screen these effects from observation in the lab frame, this screening is imperfect due to the finite size of the nucleus and relativistic electron motion.   The  portion of the nuclear EDM that survives electron screening is characterized by the Schiff moment \cite{Schiff63}.
While the most precise EDM measurement of a diamagnetic atom was performed on ${}^{199}$Hg \cite{Griffith09,Graner16}, \Ra\ is a promising isotope for an EDM search because octupole deformation of the \Ra\ nucleus enhances its EDM by over two orders of magnitude compared to ${}^{199}$Hg \cite{Auerbach96,Spevak97,Dzuba02,Dobaczewski05}.  Nuclear octupole deformation also simplifies EDM calculations for \Ra\ making them more consistent compared to ${}^{199}$Hg \cite{Ban10}.

There are two primary obstacles to achieving an EDM measurement in \Ra.  First, \Ra\ is radioactive with a 14.9 day half-life.  This makes \Ra\ atoms scarce and difficult to handle.  Second, the vapor pressure of radium is too low to perform spin precession measurements in a vapor cell.  Fortunately, the atomic structure of \Ra\ allows us to circumvent these challenges using laser cooling and trapping.

Recently, we reported the first measurement of the \Ra\ EDM  \cite{Parker15}, representing both the first EDM measurement using an atom with an octupole deformed nucleus and the first EDM measurement using laser-trapped atoms.  Although the $95\%$ confidence upper limit from this measurement is not yet competitive with that of $^{199}$Hg for limiting \textit{CP} violation in the nucleus according to current theoretical calculations \cite{Chupp15}, \Ra\ measurements have the potential to achieve rapid improvements by many orders of magnitude through a series of experimental upgrades.  In this work, we report the first such upgrade which allows us to improve the \Ra\ EDM $95\%$ confidence upper limit by a factor of 36 to $\left| d(^{225}\mathrm{Ra}) \right| \leq1.4\times10^{-23}$ $e$ cm.  The primary improvement over our previous measurement is an improved vacuum system and a new optical dipole trap (ODT) \cite{Grimm00} geometry which prolongs the atom trap lifetime and allows us to extend spin precession measurements from 2 s to 20 s.

EDMs of hadronic systems (neutrons, diamagnetic atoms and molecules) can be described as a linear combination of several potential sources of CP violation. The traditional method used to interpret EDM limits in these systems is to assume that the EDM is generated by only one source.  
This approach provides a simple way to directly interpret an EDM limit from a single system but fails to account for the scenario where an EDM arises from multiple sources of CP violation.  
Presently, experimental limits on EDMs of several systems have been reported and it is now possible to combine EDM limits from these systems to simultaneously constrain multiple sources of CP violation \cite{Chupp15}.  Within this framework, when the \Ra\ EDM limit reaches $10^{-25}$ $e$ cm, it will improve constraints on all CP violating interactions in the nuclear medium.

\begin{figure*}
\center{\includegraphics[width=0.85\linewidth]{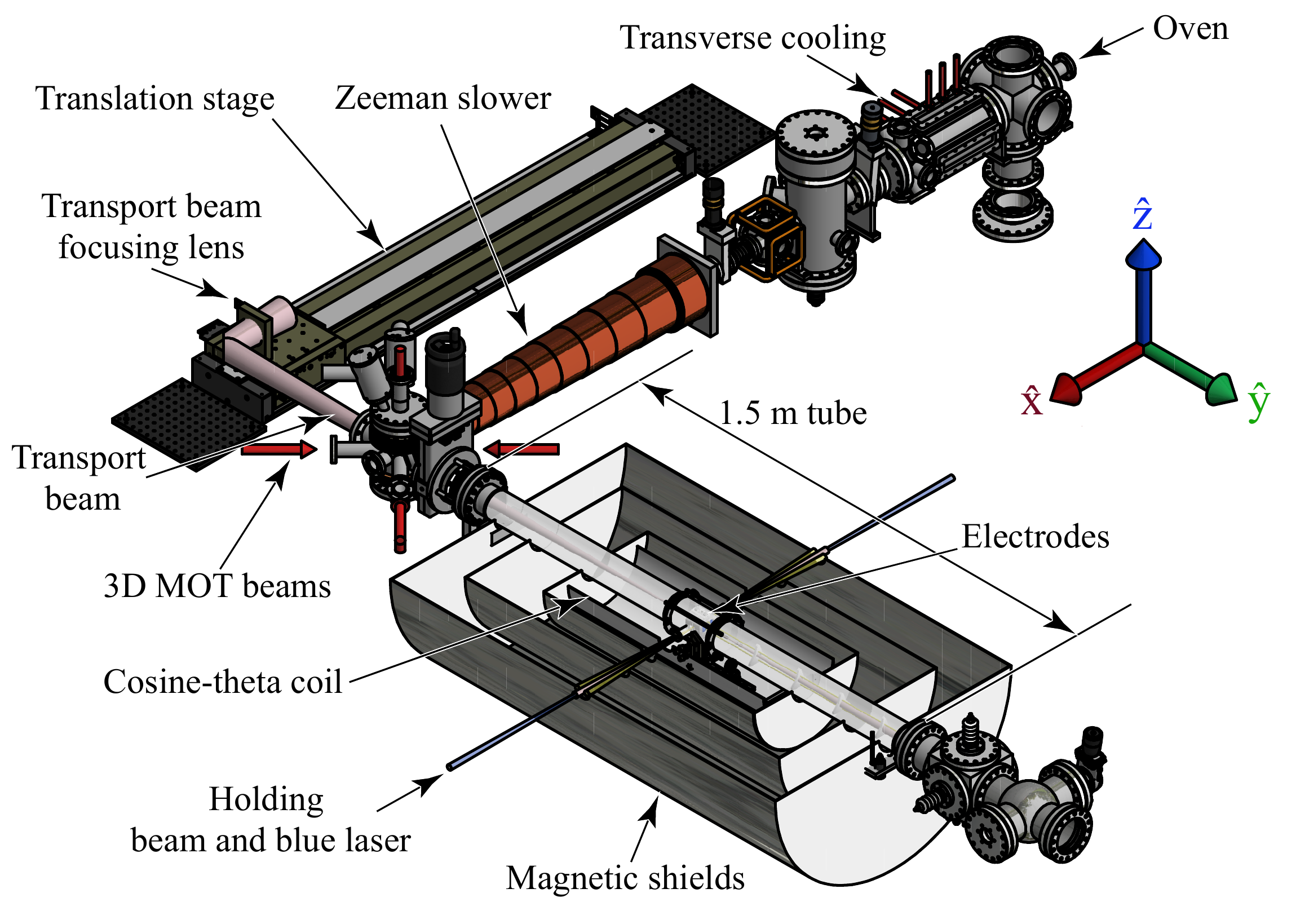}}
\caption{\label{fig:apparatus} A diagram of the experimental apparatus with important sections labeled.  Atoms exit the $\approx$500$^\circ$ C oven through a narrow nozzle where they are transversely cooled and collimated, longitudinally slowed, and trapped in a 3D MOT using laser light at 714 nm.  One out of every $10^6$ atoms exiting the oven is trapped in the MOT.  The MOT is overlapped with the focus of at 50 W 1550 nm laser which forms an ODT.  The focusing lens for the ODT is translated by $\approx$1 m to transfer the atoms between copper electrodes in the center of a glass tube with a vacuum below $10^{-11}$ torr.  We start with $\approx$10,000 \Ra\ atoms in the MOT and trap $\approx$1,000 atoms between the electrodes.  After the atoms Larmor precess for 20 s in a B-field along $\hat{z}$ we detect $\approx$500 atoms.  An E-field of 65 kV/cm is applied either parallel or anti-parallel to the B-field during spin precession.    }
\end{figure*}

\section{\label{sec:exp}Experimental Setup}

We measure the EDM of \Ra\ using a tabletop apparatus that employs widely used techniques in atomic physics \cite{Balykin84,Phillips82,Raab87,Grimm00}.  Figure \ref{fig:apparatus} shows a diagram of our apparatus.  
Section \ref{subsec:loading} introduces the relevant atomic structure that allows us to manipulate radium atoms and describes the journey the atoms take through our apparataus from when they are loaded into the oven to when they arrive at the EDM measurement region.
Here, we emphasize the changes and improvements since laser cooling and trapping of radium \cite{Guest07} and ODT transport of radium atoms \cite{Parker12} were first demonstrated.  Section \ref{subsec:measurement} describes the measurement procedure we use to interrogate the atoms after they arrive at the EDM measurement region.  Our experiment operates cyclically such that each experimental ``cycle'' lasts 100 s.   For each cycle, 60 s is spent preparing a new sample of atoms for interrogation and 40 s is spent interrogating the atoms using two separate 20 s precession measurements.  
	    
\subsection{\label{subsec:loading}Atom preparation}

\begin{figure*}
\center{\includegraphics[width=\linewidth]{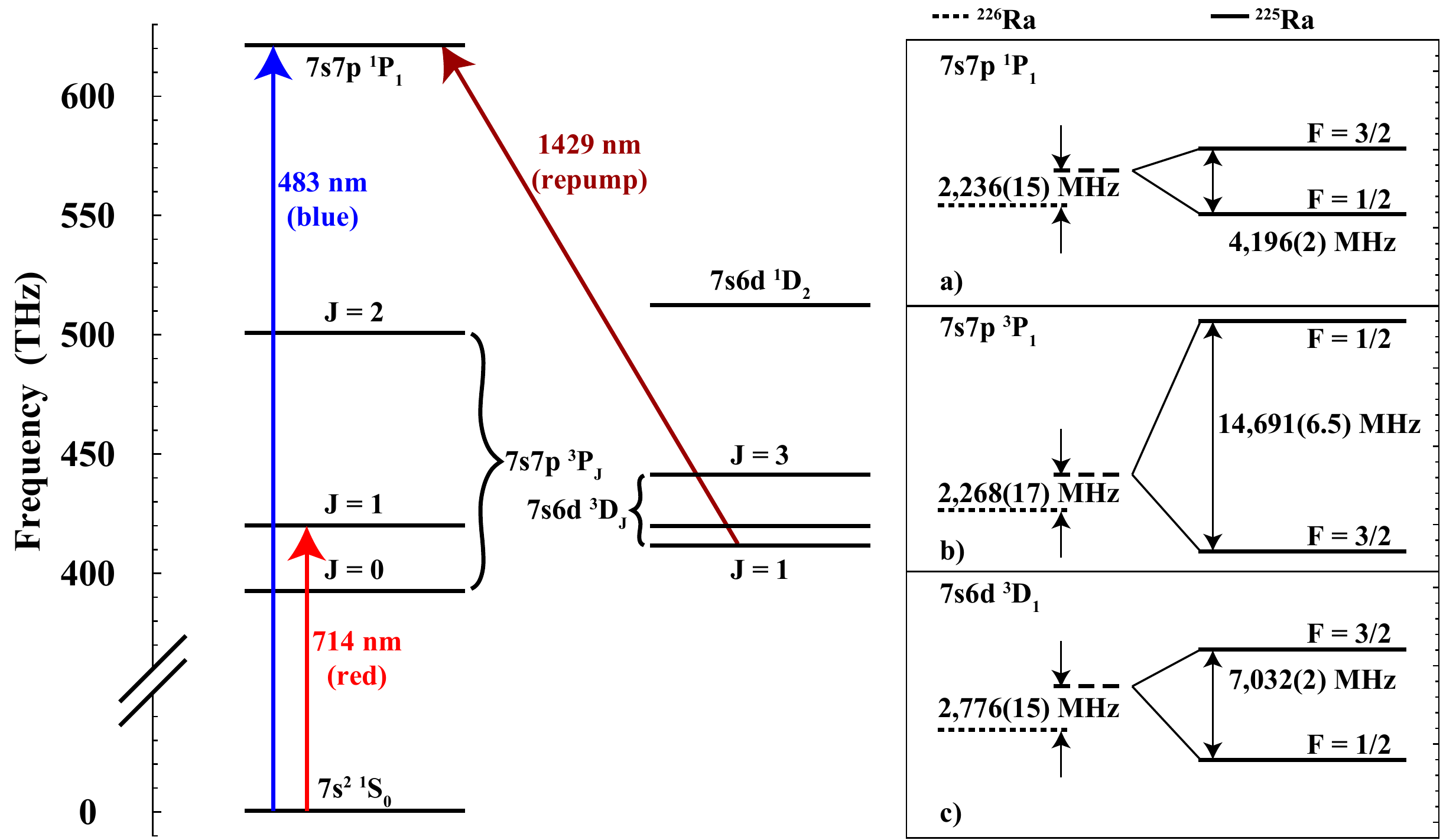}}
\caption{\label{fig:level_diagram} A level diagram including the lowest 9 energy levels of radium.  The frequency axis is plotted to scale for the upper 8 levels.  States with orbital angular momentum L=2 (D-states) are offset horizontally for clarity.  Colored arrows represent the atomic transitions we drive with laser light in our experiment.  Transitions at 714 nm, 1429 nm, and 483 nm are used for cooling/trapping, repumping, and detection, respectively.  Insets a), b), and c) show the isotope shift between \ra\ and \Ra\ and the hyperfine structure present in \Ra\ resulting from the nuclear spin $I=1/2$.  Dotted lines are the \ra\ energy levels.  
Dashed lines are offset from the \ra\ levels by the isotope shift between \Ra\ and \ra.  The isotope shift represents the frequency difference between a \ra\ state and the center of gravity of the corresponding \Ra\ hyperfine states, indicated by the solid lines.  
The upper left corner of each inset is labeled with the energy level it describes. The vertical axes in all three insets have a consistent scale such that the plotted separations accurately represent relative level separation between hyperfine states.  Frequency values in a) and c) are taken from \cite{Guest07} and values in b) are taken from \cite{Scielzo06}.}
\end{figure*}

Figure \ref{fig:level_diagram} shows the  electronic energy levels of radium and the hyperfine structure of \Ra\ for the three states that are relevant for laser cooling, trapping, and detection. In \Ra, each state with $J\neq0$ is split into two hyperfine states with $F=J\pm I$.  Here $J$ is the total \textit{electronic} angular momentum and $I=1/2$ is the nuclear spin.  The atoms are transversely cooled, slowed and trapped using laser light tuned near the ${}^1$S${}_0$ $F=1/2$ to ${}^3$P${}_1$ $F = 3/2$ (red) transition at 714 nm. Here, F is the total angular momentum including nuclear spin.  The lifetime of the red transition is 420 ns which limits the photon scattering rate for this transition to 0.38 MHz. This is almost two orders of magnitude weaker than typical primary cooling transitions used in other alkaline-earth atoms with similar electronic structure (e.g. Sr, Yb).  While experiments that use these alkaline-earth atoms leverage the dipole-allowed ${}^1$S${}_0$ to ${}^1$P${}_1$ (blue) transition,  this transition is especially cumbersome in radium because branching ratios from ${}^1$P${}_1$ to $^1$D$_2$ and metastable $^3$D$_{2,1}$ states are so large that a complicated repumping scheme is necessary both for longitudinal slowing \cite{Phillips82} and magneto optical trap (MOT) \cite{Raab87} operation.  In contrast, the red transition requires a single repump laser at 1428.6 nm which excites the $^3$D$_1$ to $^1$P$_1$ transition \cite{Guest07}.

We obtain \Ra\ from the National Isotope Development Center at Oak Ridge National Laboratory.  Our previous EDM measurement combined two separate data runs which used 3 mCi and 6 mCi shipments of \Ra\ respectively \cite{Parker15}.  In the current measurement, we use a single 9 mCi shipment, which equates to 225 ng of \Ra.  The \Ra\ is delivered to our laboratory as radium nitrate salt.  We dissolve the salt in a solution of nitric acid along with  4 $\mathrm{\mu}$Ci of \ra.  We can selectively cool and trap either isotope independently and with no isotopic contamination by changing the frequency of the lasers we use to cool and trap atoms.  We use \ra\ ($\tau_{1/2}=1,600$ y) to setup and optimize the experimental apparatus since the greater abundance of this isotope allows us to load more atoms into the oven. Unfortunately, \ra\ has zero total angular momentum in its ground state since it lacks nuclear spin and thus, cannot have an EDM in the lab frame.  The radium nitrate solution is deposited on a 2.5 cm square piece of aluminum foil and allowed to dry.  Before loading the foil into the oven, we place two 25 mg granules of metallic barium on the foil and wrap the foil around the barium.  The foil and two additional 25 mg granules of barium are then placed in the oven.  With a sufficient vapor pressure of barium, the radium nitrate and barium undergo an oxidation-reduction reaction to produce metallic radium.

The heart of the oven assembly is a titanium crucible, in which we place the radium-coated foil and barium.  Atoms enter the vacuum chamber through a nozzle with a cylindrical opening that is 8.3 cm long and 0.15 cm in diameter.  The back side of the crucible attaches to a vacuum flange via a stainless steel tube that is perforated to reduce heat conductance to the vacuum chamber.  Two radiative heaters made from tungsten filament coil surround the crucible and nozzle.  We heat the oven to between $400^\circ$ C and $520^\circ$ C.  During an EDM measurement, the \Ra\ flux degrades due to oven depletion and  radioactive decay.  We gradually increase the oven temperature to maintain a constant \Ra\ flux.  Once the oven approaches $520^\circ$ C we notice a decrease in the trap lifetime and  further increases in the oven temperature do not result in a useful increase in atom flux.
Three layers of in-vacuum passive heat shields surround the heaters inside a fourth layer of heat shielding that is water-cooled.


Transverse cooling \cite{Balykin84} of the atomic beam is achieved using 150 mW of near-resonant 714 nm laser light that is expanded to a $\approx$2 cm beam diameter ($1/e^2$) and split into two orthogonal beams propagating at near normal incidence to the atomic beam propagation direction ($\hat{x}$).  The beams make multiple bounces between two pairs of 2.5 cm by 18 cm mirrors to increase their interaction time with atoms and improve collimation by slightly decreasing their angle with respect to the atomic beam normal after each reflection.  Each dimension individually gives a gain of $\approx$9 in the MOT loading rate and they combine to give a total gain of $\approx$80.

Before we can capture atoms in a MOT, it is necessary to slow their longitudinal velocity.  We use a Zeeman slower \cite{Phillips82} operating on the red transition. 
The magnitude of deceleration is determined by the lifetime of the laser-excited state, or equivalently, the line-width of the transition.  
The lifetime of the red transition limits our 0.9 m slowing region to a capture velocity of 60 m/s. 
This velocity class represents only $0.5\%$ of the atoms exiting our oven for a typical operating temperature of 500$^\circ$ C.

Slowed atoms are trapped in a MOT which combines three orthogonal pairs of circularly polarized, counter-propagating laser beams with a quadrupole B-field to create a trapping potential.  We operate the MOT in three ``phases'' which are distinguished by the intensity and detuning of the MOT laser beams and the B-field gradient we apply.  The details of each phase have been described previously \cite{Parker12}.  During the first phase we set the beam intensity, frequency detuning, and B-field gradient to optimize loading atoms into the MOT.  After loading atoms for 50 s, we decrease laser intensity and detuning while increasing B-field gradient to compress the atoms and image their fluorescence on a CCD camera.  Finally, we further decrease laser intensity and detuning to optimize cooling and loading the atoms into an ODT.

The MOT is overlapped with the focus of a 50 W, 1550 nm laser beam (transport beam).  Having a $1/e^2$ radius of 50 $\mu$m at its focus, the transport beam forms an ODT with a depth of $\approx$400 $\mu$K.  The focus is formed using a 10 cm diameter lens with a focal length of 2 m.  Atoms are transferred to the transport beam with near unity efficiency in the final phase of the MOT.  After transfer is complete, we extinguish the MOT laser beams and translate the focusing lens by approximately one meter to transfer the atoms from the MOT chamber to the EDM measurement region. Optimum transfer efficiency occurs for a sinusoidal position versus time motion profile with a transport time of 9 s.  Previously, a 6 s transport time was used \cite{Parker15}, however, improvements to the vacuum system reduced atom loss due to background gas collisions and created a new optimum transport time.  

The measurement region is located at the center of a 1.5 m long glass tube which extends from the MOT chamber horizontally along $\hat{y}$.
Here, the atoms are compressed in a one-dimensional (1D) MOT along $\hat{y}$ and transferred to a second ODT based on a 20 W, linearly polarized, single mode 1550 nm laser with a 50 $\mu$m $1/e^2$ radius at its focus (the holding beam).  This transfer procedure has been described in detail previously \cite{Parker12}.  After the atoms are transferred into the holding beam, the transport beam is extinguished along with all 714 nm laser light and the 1D MOT coil loop circuit is electrically disconnected by a relay.

The holding beam propagates along $\hat{x}$ and is linearly polarized along $\hat{y}$.  Our previous EDM measurements \cite{Parker15} used a 10 W retro-reflected holding beam.  The interference pattern generated by this previous geometry more tightly confines the atoms along $\hat{x}$.  Tighter confinement reduces sensitivity to some systematic effects, however, this geometry also reduces the trap lifetime to 10 s due to laser induced heating.  With a non-retro-reflected 20 W holding beam, we achieve a trap lifetime of $\approx$40 s.  The increased sensitivity to systematic effects is not a concern in the current measurement.  We have re-evaluated all known systematic effects for this measurement and find them to be well below the current level of statistical sensitivity (See Sec.\ \ref{sec:systematics}.)  

\begin{figure}
\center{\includegraphics[width=\linewidth]{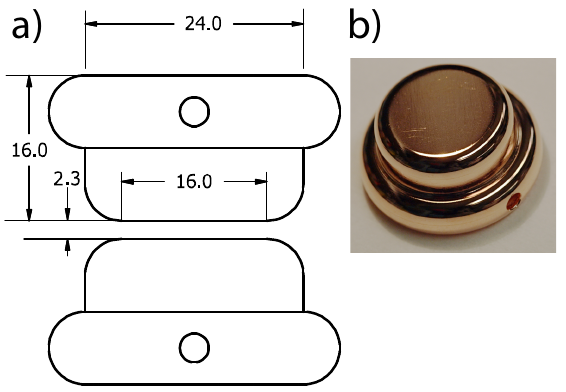}}
\caption{\label{fig:electrode}  a) A cross-sectional diagram of the copper electrodes used in the apparatus.  Measurements are labeled in mm. The solid circle represents a hole for connecting to a 3.2 mm diameter copper lead through which the electrodes are connected to the high voltage power supply.  The rounded edges have a 4 mm radius of curvature.  b) A picture of one of the copper electrodes used in this work.  }
\end{figure}

In the measurement region, button-shaped copper electrodes are located directly above and below the atoms.  The electrode surfaces facing the atoms are circular with a 1.6 cm diameter. (See Fig.\ \ref{fig:electrode} for a detailed diagram.)  Atoms are trapped near the center of the $2.3\pm0.1$ mm electrode gap.  A Macor support structure holds the electrodes in the glass vacuum chamber.  Outside the vacuum chamber, coils for the 1D MOT are wrapped around an aluminum spacer that fits snugly around the glass tube.  
A cosine-theta coil is wound around a larger cylindrical aluminum tube oriented with its symmetry axis along $\hat{y}$ to provide a uniform B-field along $\hat{z}$ with a magnitude of 2.6 $\mu$T at the location of the atoms.
 Three layers of mu-metal shielding enclose this measurement region with a shielding factor of 20,000 for B-fields along $\hat{z}$.

\subsection{\label{subsec:alignment}Field Alignment}

\begin{table}
\caption{\label{table:vectors} Measured orientation of applied fields.  Unit vectors for applied fields are expressed in terms of their Cartesian components.}
\begin{ruledtabular}
\begin{tabular}{lccc}
Vector & $\hat{x}$ & $\hat{y}$ & $\hat{z}$ \\
\hline \\
$\hat{\epsilon}_s$ & $-0.062\pm0.010$ & $-0.010\pm0.010$ & $0.998\pm0.002$ \\
$\hat{B}$ & $-0.010\pm0.061$ & $-0.068\pm0.061$ & $0.998^{+0.002}_{-0.004}$ \\
$\hat{k}$ & $1^{+0}_{-0.001}$ & $0\pm0.03$ & $0\pm0.03$ \\
$\hat{\epsilon}$ & $0\pm0.030$ & $1^{+0}_{-0.001}$ & $0\pm0.001$ \\
$\hat{b}$ & $0\pm0.001$ & $0\pm0.03$ & $1^{+0}_{-0.001}$ \\

\end{tabular}
\end{ruledtabular}
\end{table}  

Fields we apply to the atoms during precession measurements are oriented to minimize sensitivity to systematic effects.  Therefore, an accurate evaluation of their relative orientations is critical to estimating the importance of numerous systematic effects.  For the current measurement,  a rough alignment of the applied fields based on the mechanical constraints of our apparatus is sufficient to suppress systematic effects to below the current level of statistical sensitivity (See Sec. \ref{sec:systematics}).  However, it is still necessary to accurately measure the orientation of the applied fields relative to each other to reliably estimate the magnitude of these effects.  Furthermore, ongoing upgrades to our measurement technique will enable \Ra\ EDM measurements with a statistical sensitivity at the $10^{-26}$ $e$ cm level.  We show in Sec. \ref{sec:systematics} that to avoid systematic effects at this level, we must align our fields to within 0.002 rad of their design orientations.  The following section describes our evaluation of the current field orientation and the method we will use to achieve the more stringent alignment requirements of future measurements with improved statistical sensitivity.

To evaluate the present field orientation, we use gravity ($-\hat{z}$) and the projection of the glass tube in the horizontal plane ($\hat{y}$) as reference dimensions and measure the orientation of the fields relative to reference surfaces using a precision digital level with a National Institute of Standards and Technology (NIST)-traceable calibration and 350 $\mu$rad absolute accuracy. The orientation of fields in the horizontal plane is determined using mechanical tolerances of the apparatus.  We determine the direction of the following relevant fields: the applied static E-field (the given direction is for E-field applied parallel to the B-field), $\hat{\epsilon}_s$, the applied B-field, $\hat{B}$, the holding beam propagation, $\hat{k}$, the holding beam polarization, $\hat{\epsilon}$, and the holding beam B-field, $\hat{b}$.

By definition, the glass tube can only have a misalignment in the $\hat{z}$-$\hat{y}$ plane.  We measure this misalignment to be $0.01\pm0.01$ rad toward the $\hat{z}$ axis.  The Macor mount for the electrodes is constrained inside the glass tube such that the misalignment of the tube within the $\hat{z}$-$\hat{y}$ plane determines the misalignment of $\hat{\epsilon}_s$.  We image the electrodes along the glass tube axis using a CCD camera with a 1 to 1 imaging system to determine the electrode gap and similar images of the electrodes along $\hat{x}$ determine the misalignment of $\hat{\epsilon}_s$ in the $\hat{z}$-$\hat{x}$ plane from the reduced gap size. These measurements determine the direction of $\hat{\epsilon}_s$ which is included in Table \ref{table:vectors}.

Fluxgate magnetic field probes are mounted in a rectangular aluminum plate that is mechanically constrained to have an insignificant yaw angle (rotation about $\hat{z}$) relative to the glass tube.  We then determine the pitch and roll angles of the plate using a digital level.  The probes measure the magnetic field along three orthogonal axes offset 5 cm along the $\hat{z}$ axis from the location of the atoms.  Using the orientation of the mounting plate and the probes within the plate, we determine the direction of $\hat{B}$ and use the measured gradient of the $\hat{z}$ component of the B-field to estimate the uncertainty the 5 cm offset creates in our knowledge of the B-field direction at the atoms.  The final uncertainty in the direction of $\hat{B}$ is dominated by the manufacturer-specified uncertainty of the alignment of each probe within its sealed housing.

The holding beam is mechanically constrained, by aperture in the magnetic shields, to propagate along $\hat{x}$ to within $\pm0.03$ rad in any one direction.  The polarizer that determines the polarization of the holding beam is referenced to a laser table surface which is level to better than 0.001 rad such that $\hat{\epsilon}$ is along $\hat{y}$ with an uncertainty that is determined by the uncertainty in $\hat{k}$.  This gives that the polarization vector is constrained to the $\hat{x}$-$\hat{y}$ plane to within $\pm0.001$ rad since it requires coupled misalignment of the holding beam in two dimensions.

In future measurements, it is possible to align the applied fields to within of 0.002 rad of their design orientation using commercial products to evaluate the field alignment and the current capabilities of the experimental apparatus to adjust field orientations.  Using an autocollimator to optically evaluate in-vacuum surface orientations, we can determine the orientation of our electrode spacer, and thus, our applied E-field, relative to a level surface.  Using trim and gradient coils that are already implemented in our apparatus, we can then measure and adjust the orientation of the applied B-field to be parallel to the E-field using commercial 3-axis fluxgate magnetometers with sufficiently accurate orientation and orthogonality specifications for the three fluxgate magnetic field probes.  Finally, we can orient the propagation direction of our holding beam to be parallel to a surface that is normal to the applied E-field using irises with precise heights mounted to this surface. In this scheme, the E-field will define the $\hat{z}$ axis while the holding beam will define the $\hat{x}$ axis.     

\subsection{\label{subsec:measurement}Measurement Procedure}

We manipulate and detect the spin state of \Ra\ using 483 nm laser light resonant on the ${}^1$S${}_0$, $F = 1/2$ to ${}^1$P${}_1$, $F = 1/2$ (blue spin-dependent) transition. Our blue laser has circular polarization and co-propagates with the holding beam along $\hat{x}$.  
We adjust the polarization purity to be greater than $99\%$ at a polarimeter that samples the beam after it has passed through the vacuum chamber.    
The circularly polarized blue laser is used to both polarize and detect the atoms.  To polarize the atoms, we apply a 150 $\mu$s laser pulse that optically pumps the atoms such that their nuclear spins point along $\hat{x}$.  In this state, the atoms no longer scatter photons from the blue laser and so we call this the ``dark'' state.  

The applied B-field is along $\hat{z}$ so, when the polarizing pulse is over, atoms Larmor precess about the B-field.  After 1/2 of a precession period, we say that the atoms are in the ``bright'' state, because their probability to scatter a photon from the blue laser is maximal.  An atom excited by the blue laser has a 2/3 probability to decay to the bright state and a 1/3 probability to decay to the dark state, thus, we can scatter a maximum of 3 photons per atom on average.  

For atom detection, the collimated blue laser and the shadow cast by atoms scattering photons out of the laser are imaged on a CCD camera using a 300 mm focal length lens located 600 mm from the atom cloud.  To produce a ``detection image'' we activate the CCD camera while a 60 $\mu$s laser pulse is applied to the atom cloud.  For the laser intensity we use, this pulse duration optimizes the signal to noise ratio (SNR) of the atom signal and corresponds to scattering 2.1 photons per bright state atom \cite{ParkerThesis}.  The laser beam is much larger than the imaged region so in the absence of bright state atoms we get an average of 2,200 counts per pixel on the CCD camera.  Bright state atoms scatter photons out of the laser beam and the CCD image contains both the laser beam intensity profile and the shadow that the atoms cast by scattering photons out of the laser beam.  The shadow of \Ra\ atoms depletes  $\approx$100 counts from CCD pixels that correspond to the atom cloud location.

\begin{figure*}
\center{\includegraphics[width=\linewidth]{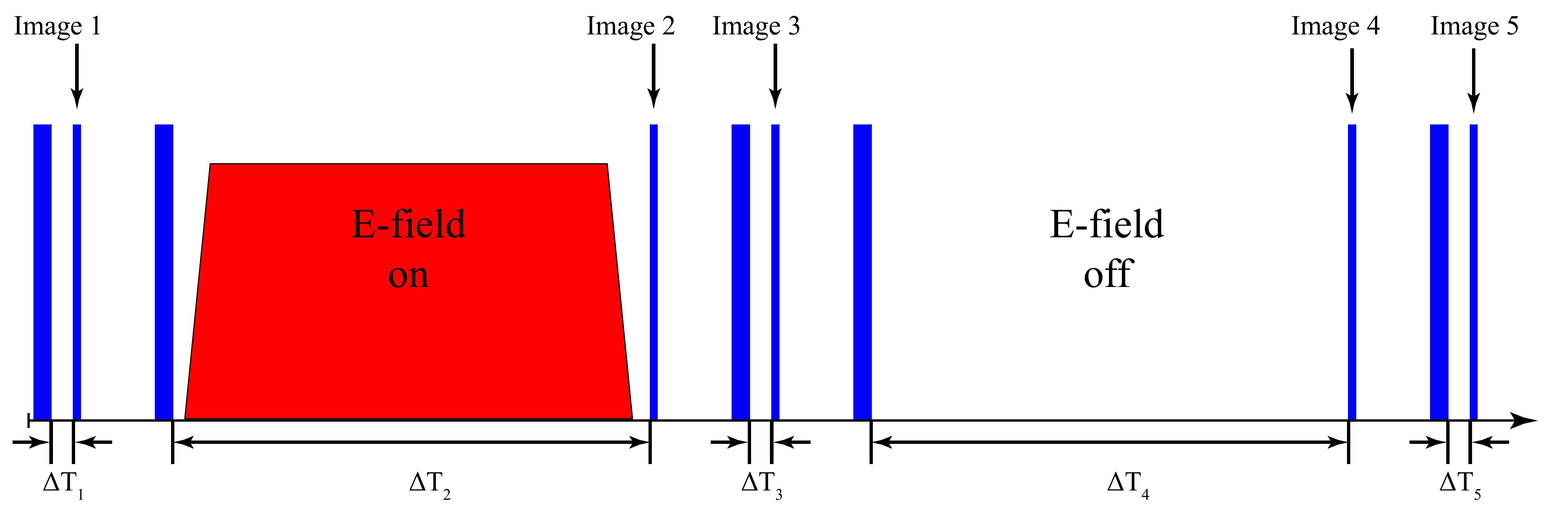}}
\caption{\label{fig:pulse_diagram} A diagram of the polarizing and detection pulse sequence used in this measurement.  Wider blue bars represent polarizing pulses and the narrow blue bars are detection pulses.  The red trapezoid designates when the E-field is ramped on. The label $\Delta T_i$ is used to denote the time between the end of the polarization pulse and the start of the detection pulse for image $i$. For images 1, 3, and 5, $\Delta T_{1,3,5}=17.4$ 
ms and for images 2 and 4 $\Delta T_{2,4}=20,000+\delta$ ms.  After an image is taken, we wait 
300 ms before applying another polarizing pulse to allow enough time for the camera to read out all captured pixels and become available for the next image.  
}
\end{figure*}

The atoms precess about the applied B-field with a period of $34.7\pm0.3$ ms.  The uncertainty in the precession period is derived from our EDM measurement, which is designed to be sensitive to relative phase accumulation and not absolute frequency.  For each experimental cycle, we take 5 detection images of the atoms.  Figure \ref{fig:pulse_diagram} illustrates the timing sequence.  
The time between the end of the $i^\mathrm{th}$ polarizing pulse and the start of the $i^\mathrm{th}$ detection pulse is $\Delta T_i$. The first detection image uses $\Delta T_1=17.4$ ms, corresponding to 1/2 of a precession period and maximal light scattered out of the detection pulse.  Images of atoms taken with $\Delta T_i=17.4$ ms determine the number of atoms in the trap.  
The second detection image is taken with $\Delta T_2=20,000+\delta$ ms where $\delta$ varies from -10 ms
to 40 ms (see Fig.\ \ref{fig:precession}).
 During precession before the second image, a  $\pm$67 kV/cm electric field is applied to the atoms by charging the top electrode to $\mp$15.5 kV and holding the bottom electrode at ground.   
Each experimental cycle, we alternate between applying an E-field that is parallel to the B-field and applying an E-field that is anti-parallel to the B-field.  The E-field on and off ramps are identical to avoid any systematic effects from the B-field that is induced by the ramp (see Sec.\ \ref{E-field_Ramping}).  The E-field is on at full strength for 19.2 s out of the total $20,000+\delta$ ms $\Delta T_2$.  
To avoid shifting and broadening the blue transition via the DC Stark effect in the strong E-field, the ramp on is initiated after the completion of the polarization pulse and the ramp off is timed such that the E-field is less that $1\%$ of its maximum value at the start of atom detection.

The third detection image is taken with $\Delta T_3=17.4$ 
ms.  This third image is used to normalize the second image since $\Delta T_3$ is chosen to place all atoms in the bright state.  The third image is a good measure of the number of atoms that were present during the second image since the lifetime of atoms in the trap ($\approx$40 s) is much longer than the time between images 2 and 3 (350 ms).  Thus, we can determine the nuclear spin population fraction directly by dividing the atom signal we extract from image 2 by that of image 3.  
The fourth and fifth detection images are taken with $\Delta T_4=20,000+\delta$ ms and $\Delta T_5=17.4$ ms, respectively.
These detection images are completely analogous to images 2 and 3 with one notable exception: no E-field is applied during $\Delta T_4$.  Data taken with no applied E-field tests for changes in precession frequency that are quadratic in E-field.  Since the applied parallel and anti-parallel E-fields are known to be identical only at the  0.7$\%$ level, any such effect could lead to a systematic effect in our EDM measurement resulting from imperfect E-field reversal (see Sec.\ \ref{Imperfect_E_reversal}).  Following the detection images, we extinguish the holding beam for 400 ms to remove any remaining atoms and then take several background images (see Sec.\ \ref{sec:results}).

\begin{figure}
\center{\includegraphics[width=\linewidth]{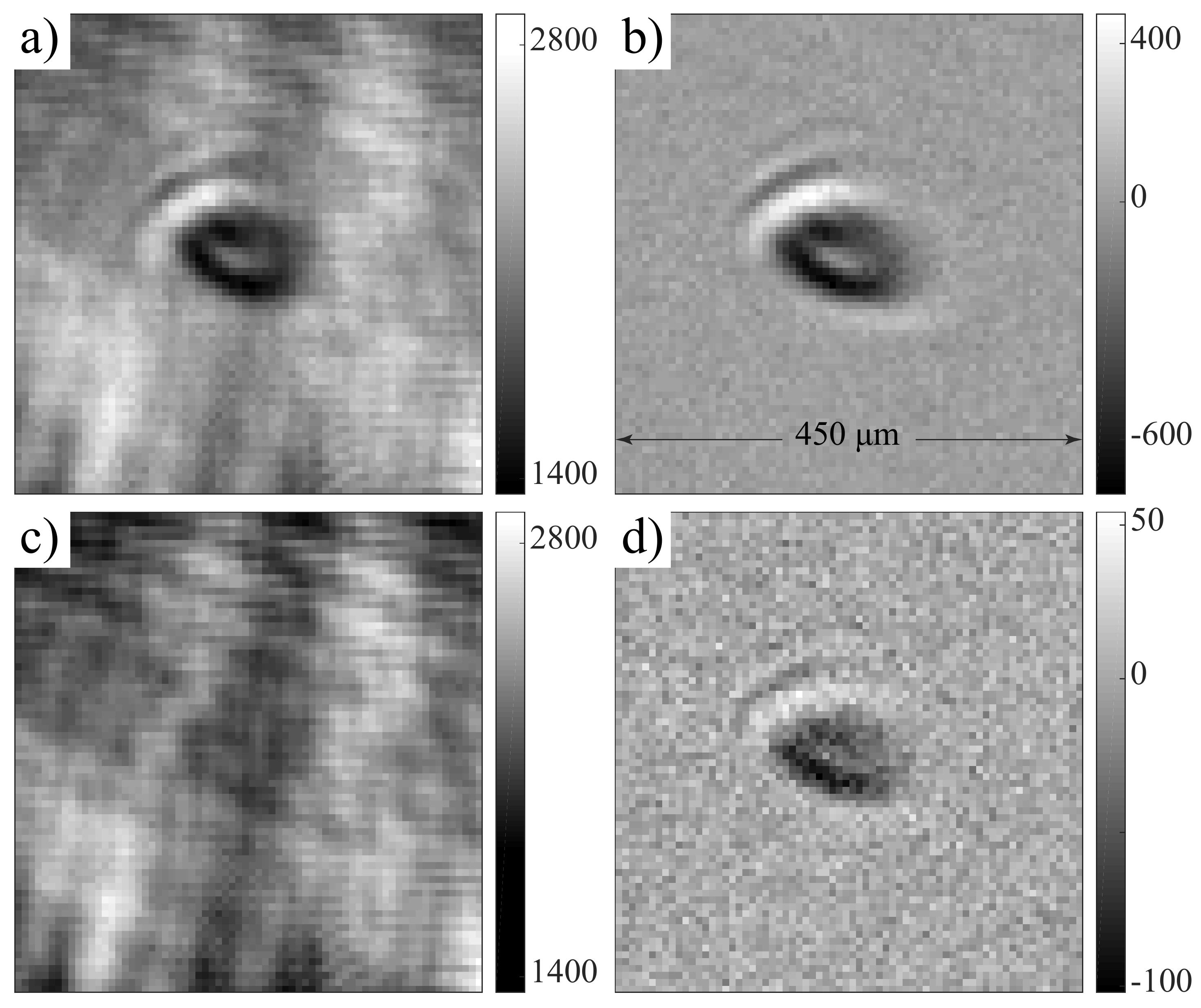}}
\caption{\label{fig:background_subtraction} A collection of 4 images that demonstrate the effectiveness of background subtraction.  a) An average of 8 \ra\ images atoms taken during the first detection image.  Periodic fluctuations are visible in the blue laser beam intensity due to interference fringes that are caused by optical elements whose surfaces create reflections that co-propagate with the input beam.  b) An average of the same images after background subtraction.  c) An average of 63 images of \Ra\ atoms taken during the first detection image.  d) An average of those same images after background subtraction.    We account for the nontrivial shape of the atom cloud image by weighting each \Ra\ image with a corresponding \ra\ image before integrating the image. }
\end{figure}

\section{\label{sec:results}Data Analysis and Results}

In the preceding section, we described the procedure we use to acquire images of the atoms and the general purpose for each image.  This section will describe how we use these images to extract the nuclear spin state of the atom cloud and how, by varying 
$\delta$, we extract the relative Larmor precession phases of the three E-field conditions using a simultaneous fit to the entire data set.

Detection images contain the intensity profile of the blue laser beam as well as the shadow of the atom cloud caused by bright state atoms scattering photons out of the laser beam.  Before we can extract information about the atoms, we need to remove the laser beam ``background'' because it contains sources of noise that are large compared to the atom signal and vary over time.  In principle, spatial fluctuations of the laser beam intensity are limited only by photon shot noise such that pixels that detect on average N photons will have fluctuations of $\sqrt{N}$.  
However, scattering from dust or other imperfections in the beam path and multiple reflection paths that interfere with the main transmitted beam all create distortions to the intensity profile of our blue laser that can change on timescales of several seconds.    

In order to properly interpret the images, we remove these distortions using  5 background images taken before the atoms arrive and 20 background images taken after they have been dropped.  The time between background images is 333 ms and the distortions change between images, but they change in a regular and repeatable manner.   
We take 25 background images in each experimental cycle which is sufficient to effectively remove background distortions.  We use a least squares fit to determine the linear combination of background images that best subtracts background distortions from each detection image \cite{Ockeloen10}, leaving behind only the atom shadow. Figure \ref{fig:background_subtraction} demonstrates the effectiveness of this technique.  The noise level we observe in background-subtracted images is only a factor of 1.2 larger than the predicted photon shot noise in a detection pulse.  

As shown in Fig.\ \ref{fig:background_subtraction}, the background-subtracted atom shadow is also distorted due to, for example, thermal effects in the imaging optics caused by the holding beam.  To compensate, we run the experimental cycle with \ra\ and average the images from each cycle to create high SNR atom images for each of the five detection images. We use these \ra\ images to weight \Ra\ images pixel for pixel before summing over all pixels to extract a number proportional to the strength of the atom signal, which we term the ``shadow magnitude''.  We create new weighting images approximately once every 24 hours to accommodate  long-term drifts in the position of the holding beam.  

Since the SNR of each individual \Ra\ shadow image is low, we average shadow magnitudes from multiple experimental cycles for each detection image. This procedure involves binning the data and determining an uncertainty for each bin based on the scatter of shadow magnitudes within each bin.  In collecting data into bins, we aim to compromise between minimizing the statistical uncertainty and accounting for slow drifts in the experimental conditions.  A detailed account of this procedure can be found in Appendix \ref{appendix:averaging}.

The probability to scatter photons out of the blue laser is maximal for the bright state and minimal for the dark state, so the shadow magnitude is proportional to the bright state population fraction.  However, the shadow magnitude is also proportional to the number of atoms in the trap during the image.  To eliminate fluctuations in the signal size due to fluctuations in atom number, we divide the averaged shadow magnitude from image 2(4) by the corresponding average shadow magnitude from image 3(5).  This value is equal to the bright state population fraction because the atoms in image 3(5) are all in the bright state.  For each E-field condition and 
$\delta$ that we measure in the experiment, we extract the bright state population fraction and its associated uncertainty by propagating the uncertainties of the averaged shadow magnitudes.   Whenever we make more than one measurement using the same 
$\delta$ and E-field condition, we calculate the weighted mean and weighted error of the mean to combine multiple measurements into a single mean value and uncertainty.

\begin{figure}
\center{\includegraphics[width=\linewidth]{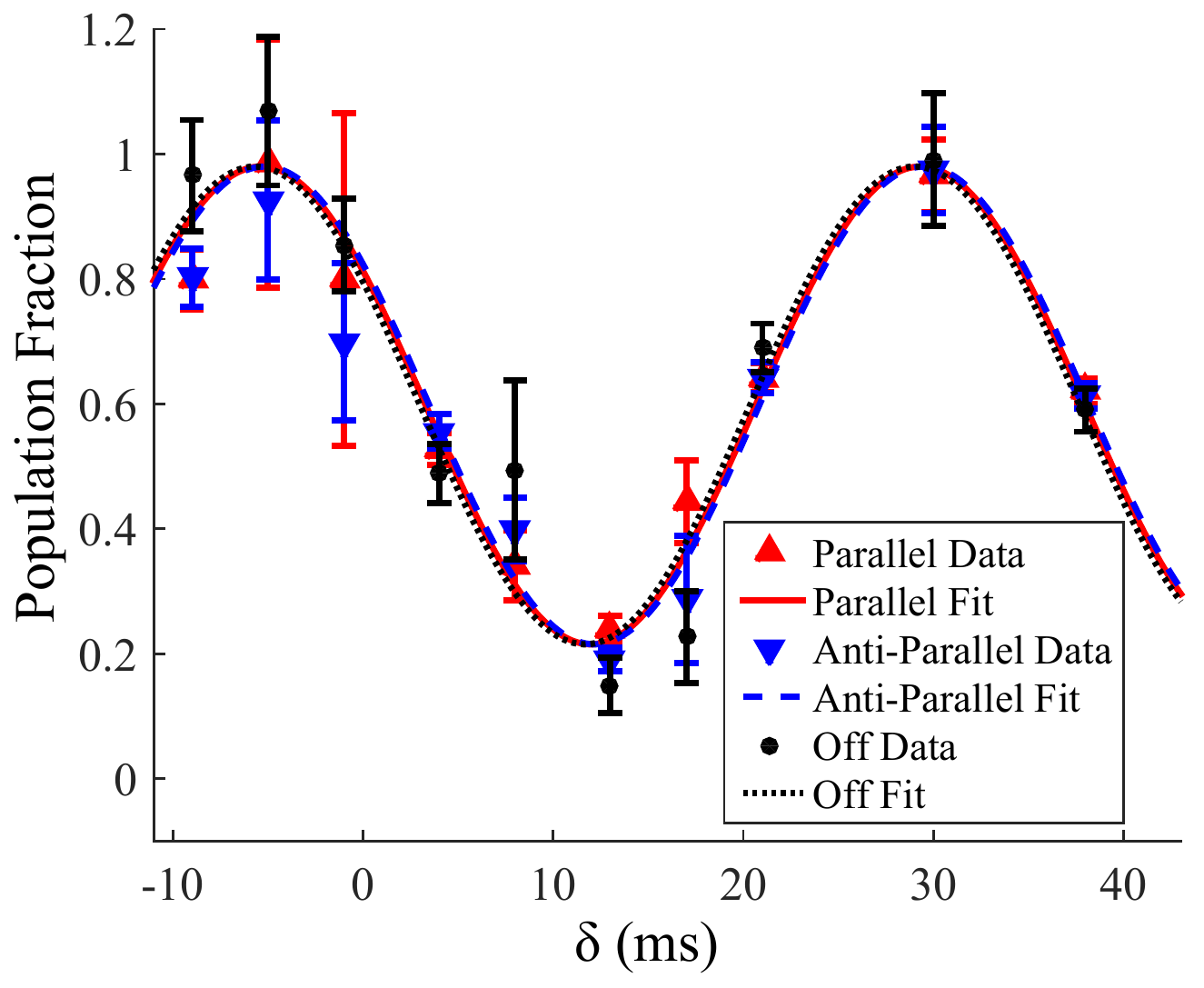}}
\caption{\label{fig:precession} Nuclear spin precession data for three E-field conditions: E-field parallel to B-field, E-field anti-parallel to B-field, and E-field off.  
The plot shows population fraction in the bright state versus $\delta$, where $\Delta T_{2,4}=20,000+\delta$ ms.
Lines represent a simultaneous fit of all the data to Eqns.\ \ref{eqn:fit_equations} using a chisquare minimization fitting program.   }
\end{figure}

The measured bright state population fraction versus 
$\delta$ is plotted in Fig.\ \ref{fig:precession} for all three E-field conditions.    We simultaneously fit this data to the three equations below using  chi-square minimization.

The equations that describe the spin precession of the three E-field conditions are:
\begin{equation}
\begin{array}{c}
y_\mathrm{off} = \frac{A}{1+P}\left[1-P\cos{(\omega \Delta T_4)}\right], \\
\\
y_\mathrm{parallel(anti\mbox{-}parallel)} = \frac{A}{1+P}\left[1- P \cos{(\omega \Delta T_2 + \theta\pm\Delta\phi/2)}\right].
\end{array}
\label{eqn:fit_equations}
\end{equation}          
Here, $A$ is a normalization constant, $P$ is the signal contrast, $\omega$ is the angular precession frequency, $\theta$ is the phase difference between E-field on and E-field off data, and $\Delta\phi$ is the phase difference between the E-field parallel and E-field anti-parallel data resulting from an atomic EDM.  We allow $A$, $P$, $\omega$, $\theta$, and $\Delta\phi$ to fit to the data.  Figure \ref{fig:precession} plots the functions from Eqn.\ \ref{eqn:fit_equations} using the best fit values for the five fit parameters. Table \ref{table:fit_parameters} lists the best values for the fit parameters and their 1-$\sigma$ standard errors.  
The uncertainties in Table \ref{table:fit_parameters} include any possible fit parameter correlations, however, the calculated covariance matrix elements show that the correlations of $\Delta\phi$, which corresponds to the EDM-induced phase difference between E-field parallel and E-field anti-parallel, with all other fit parameters are insignificant.  For example, a 1-$\sigma$ change in $A$ from its best fit value creates a 0.01-$\sigma$ change in the best fit value of $\Delta\phi$.  
Since we measure population fraction using a 
$\Delta T_{2,4}$ ($\approx$20 s) that is long compared to the variation in 
$\delta$ across the data ($\approx$50 ms), the fit finds multiple local minima for the fitted value of $\omega$.  The value we quote for $\omega$ corresponds to the global best fit and the variance is calculated as the range over which the $\chi^2$ of the fit increases by 1 globally.
For all other parameters, the variance is calculated as the inverse of the second partial derivative of the chi-square function with respect to the parameter.  With 25 degrees of freedom in the fit, we find $\chi^2/25=1.4$.

\begin{table}
\caption{\label{table:fit_parameters} Best fit values and uncertainties for fit parameters used to fit experimental data to Eqn.\ \ref{eqn:fit_equations}. Values for $\omega$ are given in rad$/$s and values for $\theta$ and $\Delta\phi$ are given in rad.}
\begin{ruledtabular}
\begin{tabular}{lll}
Parameter & Value & Uncertainty \\
\hline \\
Amplitude, A & 0.98 & 0.02 \\
Contrast, P & 0.64 & 0.02 \\
$\omega$ & 181.1 & 1.6 \\
$\theta$ & -0.070 & 0.056 \\
$\Delta\phi$ & 0.034 & 0.046 \\
\end{tabular}
\end{ruledtabular}
\end{table}

With an E-field applied parallel(anti-parallel) to the B-field, the frequency of Larmor precession is altered in the presence of a permanent electric dipole moment, $d$, according to:
\begin{equation}
\hbar\omega_{\pm} = 2\mu B \pm 2 d E.
\label{eqn:precession}
\end{equation}
Here, $\hbar$ is the Planck constant, $\omega_\pm$ is the precession frequency with the E-field applied parallel(anti-parallel), $\mu$ is the magnetic dipole moment of the \Ra\ \Ss\ ground state, and $E$ is the magnitude of the applied E-field.  For an E-field of magnitude $E$ applied over a period of time, $\tau$, Eqn.\ \ref{eqn:precession} also relates the phase difference between precession curves measured with the E-field applied parallel and anti-parallel to the permanent electric dipole moment as follows:
\begin{equation}
d = \frac{\hbar\Delta\phi}{4E\tau}.
\label{eqn:phase_to_edm}
\end{equation} 
We take $\tau$ to be the total precession time and $E$ to be the average magnitude of the applied E-field to account for the ramping on and off of the applied E-field.  Using the best fit value of $\Delta\phi$ to experimental data, we get that the electric dipole moment of \Ra\ is equal to 
$(4\pm6_\mathrm{stat}\pm0.2_\mathrm{syst})\times 10^{-24}$ $e$ cm, 
which gives a $95\%$ confidence upper limit of $\left|d({}^{225}\mathrm{Ra})\right|<1.4\times 10^{-23}$  $e$ cm.  This measurement improves upon our previous result \cite{Parker15} by a factor of 36.

\section{\label{sec:systematics}Systematic Effects}

Systematic effects mimic the signal of a true EDM by causing a true or apparent spin precession phase shift between the two conditions:  E-field parallel to B-field and E-field anti-parallel to B-field.  We have evaluated known systematic effects for the current work and the results of this analysis are included below.    
 
We use the mean value and uncertainty from our evaluation of each systematic effect to calculate a $68.3\%$ confidence upper limit and use this value as the 1-$\sigma$ uncertainty for the effect.   The uncertainties for all effects are shown in Tab.\ \ref{table:systematics}.  The total systematic uncertainty of the measurement is determined by adding the uncertainties from all individual effects in quadrature.

Certain systematic effects arise from correlations between an experimental parameter and the applied E-field.  To limit these effects, we directly measure the parameter in question throughout the data run and compare the average value during each E-field condition to determine if an E-field correlated change occurs.  Our measurement of the E-field correlated change is used to calculate the magnitude of the resulting systematic effect.

We also estimate the level at which systematic effects will limit future iterations of our experiment under two scenarios. In the first scenario ($\alpha$), we estimate the systematic sensitivity of the current experimental apparatus with the applied fields aligned to within 0.002 rad of their designed orientation.  This is possible with autocollimator-assisted installation of improved (commercially-available) fluxgate magnetometers, along with B-field trimming.  For the $\alpha$ scenario, we assume a statistical sensitivity of $10^{-26}$ $e$ cm.  In a second scenario ($\beta$), we imagine a measurement with the improved alignment from the $\alpha$ scenario, with atoms trapped in a retro-reflected holding beam geometry (as in our previous measurement \cite{Parker15}), and with an atomic magnetometer isotope co-trapped with \Ra\ in the holding beam (a co-magnetometer).  Furthermore, we assume a statistical sensitivity of $10^{-28}$ $e$ cm.  A statistical sensitivity at this level is possible by implementing the upgrades outlined in section \ref{sec:Discussion}.  
We find that the total systematic uncertainty of the current measurement is $2\times10^{-25}$ $e$ cm. For future measurements,  we estimate that in the $\alpha$ scenario we will achieve a total systematic uncertainty of $5\times10^{-27}$ $e$ cm and in the $\beta$ scenario this estimate reduces to  $4\times10^{-29}$ $e$ cm.  

\subsection{\label{Imperfect_E_reversal}E-squared effects}

Any effect that produces a phase shift proportional to the square of the applied E-field cannot result from an EDM but still can produce a systematic effect if the E-field reversal is imperfect.  In this measurement, we determine that the magnitude of the parallel and anti-parallel E-fields are matched to within $0.7\%$ using a calibrated high voltage (HV) divider.  We make spin precession measurements with no E-field applied to place an experimental limit on any spin precession phase shift proportional to the square of the applied E-field.  Using our measurement of $\theta$, the differential phase acquired during spin precession between E-field off data and E-field on data, and an E-field imbalance of $0.7\%$, we obtain a 1-$\sigma$ uncertainty of $1\times10^{-25}$ $e$ cm for this effect in the current measurement.

We do not predict that we will see phase shifts proportional to the square of the E-field even in the $\beta$ scenario.  In the current experimental procedure, we estimate that we would need a statistical sensitivity of $10^{-25}$ $e$ cm to measure known sources of phase shifts proportional to the square of the E-field.  The dominant effect arises from E-field gradients that push the atoms in opposite directions for the two E-field polarities combined with a B-field gradient that creates a position dependent precession frequency. 
Future implementations of the experiment will use a retro-reflected holding beam which confines the atoms more tightly and reduces the statistical sensitivity needed to observe this effect to $10^{-30}$ $e$ cm.

It is important to note that as the statistical sensitivity of the experiment improves, measurements of this systematic will improve in lockstep.  Measurements of phase shifts proportional to the square of the applied E-field will always have the same statistical sensitivity as the associated EDM measurement.  This systematic effect, however, is reduced by the magnitude imbalance of the applied E-fields.  For example, the current measurement has a statistical sensitivity at the $10^{-23}$ $e$ cm level and we balance the E-fields to $0.7\%$, enabling us to evaluate this effect with an uncertainty of $10^{-25}$ $e$ cm.  In the $\alpha$ and $\beta$ scenarios, we will be able to measure this effect with a precision of $7\times10^{-29}$ and $7\times10^{-31}$ $e$ cm, respectively,  using this same technique.   

\subsection{\label{B-field_Correlations}B-field correlations}
The phase acquired during spin precession is directly proportional to the applied B-field and any change of the B-field during spin precession that correlates with the applied E-field will produce a false EDM signal.  This false EDM signal is given by:
\begin{equation}
d_\mathrm{false}=\frac{\mu\Delta B}{E},
\label{eqn:B_correlation}
\end{equation}
where  $\Delta B$ is the measured difference in B-field between when the E-field applied parallel and when the E-field is applied anti-parallel.  We place three orthogonal fluxgate magnetometer probes offset 5 cm along $\hat{z}$ from the atoms and continuously measure the B-field during our EDM measurement.  We have constrained the gradient in the B-field magnitude along $\hat{z}$ to be less than $0.1\%$/cm by measuring the B-field magnitude at three locations along the $\hat{z}$-axis within the cosine-theta coil using a rubidium magnetometer.  Taking into account that this effect is only sensitive to \textit{changes} in the B-field that correlate with changes in the applied E-field, the offset of the fluxgate does not significantly contribute to our measurement of this systematic (i.e.\ for a measured E-field correlated change in B-field at the fluxgate, we need to allow for the possibility that the correlated change at the atoms is greater than the measured value by at most $1\%$ due to the 5 cm offset of the probe from the atoms).  
To achieve a precise measurement of $\Delta B$ we measure the  magnetic field during $\Delta T_2$ throughout the entire data run and take the difference between pairs of adjacent experimental cycles with opposite E-field polarities.  We average all of these differential measurement to reduce our sensitivity to magnetic field noise on timescales longer than 200 s and get that $\Delta B=-0.3\pm0.5$ pT. 
This gives a 1-$\sigma$ uncertainty of $1\times10^{-25}$ $e$ cm for this systematic effect. 

For future measurements, we have dramatically decreased B-field measurement noise by low-pass filtering our fluxgate signals.  We now achieve the manufacturer specified measurement noise of  6 pT for an integration time of 1 s.
In the $\alpha$ scenario, we estimate a 15 day measurement time where for 19.8 s of each 100 s cycle we can compare the measured B-field from the three E-field conditions to determine if an E-field correlated change in B-field occurs.  In this scenario we predict an uncertainty of $5\times10^{-27}$ $e$ cm for our measurement of this systematic effect 

In the $\beta$ scenario, a co-magnetometer can be used to measure the B-field at the atoms and with greater sensitivity.  
$^{171}$Yb or $^{199}$Hg may be attractive co-magnetometer species because they have similar atomic structure to \Ra\ and are predicted to be much less sensitive to CP violating effects in the nuclear medium \cite{Dzuba07,Dzuba02}.  
The spin precession measurement can proceed exactly as in \Ra\ and the statistical sensitivity will improve proportional to the square root of the number of atoms.  Since the co-magnetometer atom will be a stable isotope, we conservatively predict that we will be able to trap a factor of ten more magnetometer atoms than \Ra\ atoms. This leads to a measurement of the E-field correlated change in B-field that is a factor of 3 more precise than the EDM measurement or an uncertainty of $3\times10^{-29}$ $e$ cm for this systematic effect.

\subsection{\label{Blue_Frequency_Correlations}Blue laser frequency correlations with E-field}

 Frequency fluctuations of the 483 nm absorption imaging laser that are correlated with the applied E-field could create an EDM-like signal.  The current measurement is insensitive to any such correlation since we split the data between parts of the precession curve with opposite slopes.  However, this will still lead to a differential amplitude between the two E-field polarities, which can create an EDM systematic through correlations between the fit parameters $A$ and $\Delta\phi$.  
 
We measure the E-field correlated frequency fluctuations of our 483 nm laser by continuously monitoring light transmission through a passive optical cavity that serves as a frequency reference using a battery powered phototdiode.  We measure the fractional amplitude of E-field correlated transmission fluctuations, $\Delta A_\mathrm{cav}$ to be $-75\pm80$ ppm which gives a $68.3\%$ confidence upper limit of 115 ppm.  To connect cavity transmission fluctuations to atomic scattering fluctuations, we use a peak normalized Lorentzian function to describe the cavity and atomic resonances as a function of frequency.

To proceed with the most conservative estimation of an EDM systematic effect, we assume that the light is exactly centered on the cavity resonance, where a measurement of intensity fluctuations is insensitive to frequency fluctuations and that the laser light is detuned from atomic resonance by $\Gamma_\mathrm{Ra}/(2\sqrt{3})$, where $\Gamma_\mathrm{Ra}$ is the full width at half maximum (FWHM) of the blue transition.  At this detuning, the amplitude of light scattered by atoms is most sensitive to laser frequency.  For these assumptions, it can be shown that the fractional fluctuations in atomic scattering amplitude, $\Delta A_\mathrm{Ra}$, is related to $\Delta A_\mathrm{cav}$ by
\begin{equation}
\Delta A_\mathrm{Ra} = \frac{3\sqrt{3}}{8}\frac{\Gamma_\mathrm{cav}}{\Gamma_\mathrm{Ra}}\sqrt{\Delta A_\mathrm{cav}},
\label{eqn:scattering_fluctuations}
\end{equation}
where $\Gamma_\mathrm{cav}$ is the FWHM of the cavity resonance and we assume $\left|\Delta A_\mathrm{cav}\right|\ll1$.

Then the false phase shift between E-field polarities caused by correlations between $A$ and $\Delta\phi$ is given by,
\begin{equation}
\Delta\phi_\mathrm{false}=\Delta A_\mathrm{Ra}\frac{\rho_{A,\Delta\phi}}{\rho_{A, A}}.
\label{eqn:fake_phase_blue_frequency}
\end{equation}
Here, $\rho_{A,\Delta\phi}$ is the covariance matrix element connecting fit parameters $A$ and $\Delta\phi$ and $\rho_{A,A}$ is the variance of $A$.  Finally, we can calculate the false EDM signal with the added suppression of 0.09 because we detect atoms at least 390 ms after the E-field has been ramped down with a $1/e$ time of 160 ms.  We calculate the systematic effect of blue laser frequency fluctuations to have a 1-$\sigma$ uncertainty of $4\times10^{-28}$ $e$ cm in our current measurement.  

In both the $\alpha$ and $\beta$ scenarios, we can improve our measurement of blue laser frequency fluctuations by locking to the side of the cavity resonance. 
This changes the dependence of $\Delta A_\mathrm{Ra}$ on $\Delta A_\mathrm{cav}$ from Eqn.\ \ref{eqn:scattering_fluctuations} to,
\begin{equation}
\Delta A_\mathrm{Ra} = \frac{\Gamma_\mathrm{cav}}{\Gamma_\mathrm{Ra}}\Delta A_\mathrm{cav}.
\label{eqn:scattering_fluctuations_new}
\end{equation}
This gives that the upper limit on a false EDM signal due to E-field correlated changes in blue laser frequency would be $8\times10^{-30}$ $e$ cm. 

\subsection{\label{Blue_Power_Correlations}Blue laser power correlations with E-field}

As part of our measurement, we image each blue laser detection pulse on a CCD camera.  For the EDM measurement, we subtract the ``background'' signal and look only at the ``shadow'' of the atoms that they create by scattering photons out of the blue laser pulse.  However, the background is a direct measure of the blue laser power used to image the atoms.  For adjacent experimental cycles, we compare the backgrounds subtracted from detection image 2 to obtain the difference in blue laser power between images taken directly after the application of a parallel and anti-parallel E-field. Combining all pairwise measurements into a weighted mean and error, we get that the 1-$\sigma$ uncertainty in correlated power fluctuations is $0.2\%$.  For the most conservative systematic estimation we assume that we are far below saturation such that fluctuations in laser power are directly proportional to fluctuations in atomic scattering.  Then using the analysis of Section \ref{Blue_Frequency_Correlations} to connect fluctuations in atomic photon scattering to a false EDM signal, we limit systematic effects from blue laser power correlations with E-field to a 1-$\sigma$ uncertainty of $7\times10^{-28}$ $e$ cm.

For this measurement, the standard deviation of fractional pairwise power differences is $3\%$.  However, the fundamental photon shot noise of the detection pulse is $0.2\%$.  By implementing intensity feedback on the acousto-optic modulator that creates the blue laser pules for atom detection, we can achieve shot-noise-limited uniformity in laser pulse intensity.  The statistical sensitivity of our blue laser power correlation measurement will improve proportional to the improvement in pulse intensity variation.  Furthermore, the \emph{sensitivity} of the EDM measurement can be additionally suppressed by normalizing the atom absorption signal by the laser pulse power.  Thus, we will be able to improve our evaluation of this systematic to $1\times10^{-31}$ $e$ cm.

\subsection{\label{Holding_power_correlation}Holding ODT power correlations}

A correlation between the holding beam power and the applied E-field could lead to a systematic effect in our measurement.
For example, residual circular polarization in the holding beam causes a vector AC stark shift of the ground state that shifts the Larmor precession frequency proportional to the holding beam power.    
We limit the contribution of this effect using measured correlations between holding beam power and applied E-field along with a calculation of the shift in \Ra\ that is similar to a calculation reported for $^{199}$Hg \cite{Romalis99}.

For diamagnetic atoms having nuclear spin, such as \Ra\ and $^{199}$Hg, the vector AC Stark shift is due to hyperfine structure in excited states.  Since the dominant contributions arise from transitions to the \Pt{1} and \Ps\ states, we sum over the AC Stark shifts from these transitions to obtain the differential shift that would occur between the ground state Zeeman sub-levels in a circularly polarized laser beam, $\nu_v$.  Using the parameters of the current 20 W, single pass holding beam, we calculate $\nu_v = 50$ Hz.  Since linearly polarized light cannot cause a vector shift, it is suppressed in our experiment from the linear polarization of the holding beam.  The holding beam passes through a calcite polarizer with a 100,000:1 extinction ratio before entering the vacuum chamber and we measure the polarization purity to be $>99.9\%$ after passing through the vacuum chamber, which implies a polarization at the atoms of better than 99\%, with a high level of confidence.  Also, since the vector shift is directed along the laser’s propagation direction, which is perpendicular to the electrodes, there is a further suppression.
The final shift can be expressed as
\begin{equation}
\Delta\nu_{m_F}=m_F\nu_V\left(|\varepsilon_L|^2-|\varepsilon_R|^2\right)\cos(\theta_{BH}),
\label{eqn:vector_shift_experiment}
\end{equation}
where $\theta_{BH}$ is the angle between the holding beam and the applied B-field and the coefficients $\varepsilon_R$ and $\varepsilon_L$ represent the right- and left-hand circular components of the holding beam polarization.
This effect is suppressed by two orders of magnitude since the holding beam polarization is over $99\%$ linear.  Furthermore, the measured field alignment gives that the $68.3\%$ confidence upper limit for $\cos(\theta_{BH})$ is 0.1.  The resulting false EDM signal is given by   
\begin{equation}
d_\mathrm{false}=\Delta\nu_{1/2}\frac{h}{2E}\frac{\Delta P}{P_0},
\label{eqn:holding_correlation}
\end{equation}
where $\Delta P/P_0$ is the fractional difference in holding beam power between applying an E-field parallel and anti-parallel to the applied B-field.  By monitoring the holding beam during E-field applications we measure no correlation between holding beam power and E-field and limit $\Delta P/P_0$ to a $68.3\%$ confidence upper limit of of $8\times10^{-5}$.  This corresponds to a 1-$\sigma$ uncertainty of $6\times 10^{-26}$ $e$ cm for the associated systematic effect on our measurement.

As the statistical sensitivity improves to the level where these effects may become detectable, we can limit this systematic in a model-independent way by directly measuring Larmor precession frequency shifts as a function of holding beam power.  With the current apparatus we can change the power of the holding beam between 20 W ($P_0$), and 10 W ($0.5P_0$) without observing a significant change in atom number.  In addition to the three field conditions we currently measure, we will add three additional field conditions corresponding to the three E-field conditions at half the holding beam power.  To extract the power shift, we will only use data with the E-field off, however, we will see in Section \ref{Stark_Interference}  that we will need to compare spin precession at two holding beam powers with the E-field on to limit Stark interference.

The power shift is related to $\phi_\mathrm{H/L}$, the measured phase shift between precession at $P_0$ and at $0.5P_0$ by 
\begin{equation}
\Delta\nu_{1/2}=\Delta\phi_\mathrm{H/L}/(\pi\tau).  
\label{eqn:vector_shift_from_phase_shift}
\end{equation}
We can use the statistical uncertainty of our EDM measurement to estimate the statistical uncertainty of $\phi_\mathrm{H/L}$.  First, we use Eqn.\ \ref{eqn:phase_to_edm} to relate the statistical uncertainty in an EDM measurement to that of a phase shift measurement.  The EDM measurement is designed to be sensitive to the E-field on conditions and we must increase the statistical uncertainty of $\phi_\mathrm{H/L}$ by a factor of $\sqrt{2}$ since there are twice as many E-field on measurement compared to E-field off measurements.  Then we use Eqn.\ \ref{eqn:holding_correlation} to relate an uncertainty in phase to an uncertainty in the false EDM signal caused by this effect.  We get that 
\begin{equation}
\Delta d_\mathrm{false}=4\sqrt{2}d_\mathrm{stat}\frac{\Delta P}{P_0},
\label{eqn:holding_correlation_projection}
\end{equation}
where $d_\mathrm{stat}$ is the statistical uncertainty of the EDM measurement, and $\Delta$$d_\mathrm{false}$ is the statistical uncertainty in our evaluation of this systematic effect.  Thus, we get that in the $\alpha$ and $\beta$ scenarios, we can evaluate this systematic with an uncertainty of $9\times10^{-30}$ and $9\times10^{-32}$ $e$ cm, respectively.

\subsection{\label{Leakage_current}Leakage Current}

Leakage current creates a false EDM signal from the induced B-field created by leakage electrons as they travel between electrodes.  Since leakage current changes sign with the applied E-field and also grows in magnitude with increasing E-field, this can be a particularly troublesome systematic.  The two most troublesome paths for leakage electrons are through the vacuum between the electrodes and along the inner surface of the Macor electrode spacer.  We use simple models to limit the effect leakage current can have in either path.  

For the current measurement, the most menacing path would be emission from one electrode such that the electrons pass close by the atom cloud but not through it.  Electrons emitted from an electrode will be accelerated by the E-field and travel in the direction of the applied E-field which induces a B-field, $B_\mathrm{ind}$, that is perpendicular to the velocity of the electrons. The dominant effect of $B_\mathrm{ind}$ is to alter the Larmor precession frequency due to misalignment between the applied E- and B-fields. We treat the electrons as an infinite wire and assume they pass within 50 $\mu$m of all atoms, which is roughly the radius of the atom cloud.  Then electrons traveling along this path give a false EDM signal that is given by
\begin{equation}
d_\mathrm{false}=\frac{\mu \mathbf{B}_\mathrm{ind}}{E}\cdot\hat{B} = \frac{\mu}{E}\frac{\mu_0 I}{2\pi r}\sin{\theta_{EB}}.
\label{eqn:leakage_systematic} 
\end{equation}
Here, $I$ is the leakage current, $\mu_0$ is the vacuum permeability, $r$ is the distance of closest approach for the electron beam, and $\theta_{EB}$ is the angle between the applied E-field and B-field.  We calculate the $68.3\%$ upper limit for $\theta_{EB}$ using the measured field orientations and their associated uncertainties to get that $\theta_{EB}\leq0.1$ rad.  The leakage current monitor during this data run measured a leakage current consistent with zero and having a $68.3\%$ upper limit of 2 pA.  Using Eqn.\ \ref{eqn:leakage_systematic}, we get a 1-$\sigma$ uncertainty for this systematic effect of $3\times10^{-28}$ $e$ cm.  

The next most important possible path for the electrons would be if they travel along the inner surface of the Macor electrode spacer, making full or partial loops along the way which induces a B-field having a component along the $\hat{B}$ direction.  This could result from (e.g.) impurities in the spacer or the irregular shape of the spacer.  To put a conservative limit on this effect, we determine the loops or partial loops that induce the largest B-field along $\hat{z}$ and imagine that the leakage current travels along this path.  Moreover, we assume that this path reverses perfectly under E-field reversal.  Because the spacer has large gaps machined into it for optical access, a solenoidal path is impossible, leaving the most important path to be a loop in the Macor near the surface where the electrode makes contact with the spacer.  With a leakage current of 2 pA, each such loop generates a systematic shift at the level of $4.5\times10^{-29}$ $e$ cm.  Even in the $\alpha$ scenario, where the electrode alignment has been corrected, this second possibility cannot be excluded, and so we use a two-loop path as an upper limit for that case, giving $9\times10^{-29}$ $e$ cm.

In the $\beta$ scenario, direct B-field measurements with a co-magnetometer would eliminate the need to individually consider the effect of leakage current because this systematic effect would be incorporated into the E-field correlated change in B-field systematic discussed in Section \ref{B-field_Correlations}.

\subsection{\label{Exv}$\mathbf{E}\times \mathbf{v}$ effects}

Atoms traveling with velocity $\mathbf{v}$ in a non-zero E-field, $\mathbf{E}$, will experience a B-field $\mathbf{B}_\mathrm{motion}$ that is equal to 
\begin{equation}
\mathbf{B}_\mathrm{motion}=\gamma\left(\frac{\mathbf{v}}{c^2}\times\mathbf{E}\right).
\label{eqn:motional_B-field}
\end{equation}
Here, $\gamma$ is approximately equal to one for the non-relativistic velocities of our atoms.  
We consider atoms at the Doppler cooling limit of 9 $\mu$K for the \Ss\ to \Pt{1} transition which have a root mean square (RMS) velocity of $v_D=0.022$ m/s in the holding beam trap if we assume harmonic motion.   Then $\left|\mathbf{B}_\mathrm{motion}\right| = 1.6\times10^{-12}$ T.  However, motion of the atoms in any one dimension is periodic with a period given by $\tau_\mathrm{trap}=2\pi/\omega_\mathrm{trap}$, where $\omega_\mathrm{trap}$ is the trap frequency.  Thus, the effect of the B-field induced during the first half of $\tau_\mathrm{trap}$ is exactly canceled by the equal and opposite B-field induced during the second half.  To put the most conservative limit possible on this effect, we imagine a synchronization between Larmor precession and atom motion such that the atom motion is identical during every precession measurement such that there is a maximal un-canceled B-field from the final $\tau_\mathrm{trap}/2$ of the total precession time $\tau$.  The induced B-field is perpendicular to $\hat{\epsilon}_s$ and it alters the precession frequency due to misalignment between $\hat{\epsilon}_s$ and $\hat{B}$.  The false EDM signal from this effect is then given by
\begin{equation}
d_\mathrm{false}=\frac{\mu\left|\mathbf{B}_\mathrm{motion}\right|}{E}\frac{\tau_\mathrm{trap}}{2\tau}\sin{\theta_{EB}},
\label{eqn:motional_B-field_systematic}
\end{equation}
We use the trap frequency for the weakly confined dimension of the holding beam ($\omega_{\mathrm{trap},\hat{x}}=2\pi\times4.25$ rad/s) to obtain the maximal possible effect.  This gives a 1-$\sigma$ uncertainty of $4\times10^{-28}$ $e$ cm for this systematic effect. In the $\alpha$ scenario, this reduces to $7\times10^{-30}$ $e$ cm due to better field alignment.
For the $\beta$ scenario, this effect is accounted for by the E-field correlated changed in B-field from Section \ref{B-field_Correlations} because the co-magnetometer atoms measure the field directly at the location of the \Ra atoms.

\subsection{\label{E-field_Ramping}E-field Ramping}

The ramp up and down of the E-field during spin precession will induce magnetic fields, which can in principle lead to a phase shift proportional to the applied E-field.  This effect cancels perfectly in the event that the ramp up and down are temporally symmetric.  In our experiment, the ramps are computer controlled by an arbitrary waveform generator to ensure the time reversal symmetry of ramping on and off.  Nevertheless, we estimate the maximum possible magnitude for this effect by considering the differential phase shift between the two E-field polarities from just the E-field ramp on without considering any cancellation of the effect from the ramp off.

There are two effects that lead to induced B-fields that are proportional to the applied E-field:
\begin{equation}
\begin{array}{c}
\mathbf{B}_\mathrm{ind} = \mathbf{B}_\mathrm{cur}+\mathbf{B}_{dE/dt},\quad \mathrm{where} \\
\\
\mathbf{B}_\mathrm{cur}=\frac{\mu_0}{4\pi}\int_C\frac{I d\mathbf{l}\times\mathbf{r}'}{\left|\mathbf{r}'\right|^3}, \quad \mathrm{and} \\
\\
\oint_{\partial\Sigma'} \mathbf{B}_\mathrm{dE/dt}\cdot d\mathbf{l}'=\mu_0\varepsilon_0\frac{d}{dt}\iint_{\Sigma'}\mathbf{E}\cdot d\mathbf{S}'.
\end{array}
\label{eqn:ramp_induced_B}
\end{equation}
Here, $\mathbf{B}_\mathrm{cur}$ is the B-field that is induced by the current, $I$, traveling through the copper lead to the electrode along path $C$ during the field ramp.  Also, $\mathbf{B}_{dE/dt}$ is the B-field that is induced on the edge of a surface, $\Sigma$, due to the changing E-field flux through the surface.  For simplicity, we assume a linear ramp and model the copper lead as a half infinite wire along $\hat{y}$ that is offset from the atoms by 13.15 mm (see Fig.\ \ref{fig:electrode}).  We consider an increasing E-field that is uniform between the electrodes and zero everywhere else and consider atoms displaced horizontally from the electrode center to the edge of the electrode, where the induced B-field is strongest.  We imagine a displacement along $\hat{y}$ such that $\mathbf{B}_\mathrm{cur}$ and $\mathbf{B}_{dE/dt}$ are in the same direction and consider the effect of $\mathbf{B}_\mathrm{ind}$ on the precession frequency due to misalignment between the applied B- and E-fields.  The observed phase shift between the two E-field polarities due to this effect is then, 
\begin{equation}
\Delta\phi_\mathrm{false}=2\pi t_0\Delta\nu_\mathrm{false}=2\pi t_04\mu |B_\mathrm{ind}|\sin{\left(\theta_{EB}\right)}/h,
\label{eqn:ramp_B-field}
\end{equation}
where $t_0$ is the duration of the ramp.  Using Eqn.\ \ref{eqn:phase_to_edm} we calculate the 1-$\sigma$ uncertainty in the corresponding false EDM signal to be $9\times10^{-28}$ $e$ cm.  In the $\alpha$ scenario, this reduces to $2\times10^{-29}$ $e$ cm due to better field alignment.  Since this  effect arises from an E-field correlated change in B-field, in the $\beta$ scenario the co-magnetometer atoms would directly evaluate this effect as a contribution to the systematic discussed in Section \ref{B-field_Correlations}.

\subsection{\label{Stark_Interference}Stark Interference}

Stark interference is an effect, third order in perturbation theory, that allows for interaction between an applied DC electric field and the AC electromagnetic field of an intense laser beam \cite{Romalis99}.  It results in a vector shift of the ground state which is linear in the DC electric field and also linear in the intensity of the laser.  Since the effect is linear in the electric field, it can mimic an EDM, and so we expect to observe a false EDM signal caused by the holding beam.  Fortunately, Stark interference can be distinguished from a true EDM by observing its dependence on the holding beam power, since only a false EDM will change in size as a function of laser power.  Here we estimate the expected magnitude of this effect.

The vector dependence of Stark interference is most simply stated if we write the frequency shift between ground state Zeeman sub-levels as the sum of two components:
\begin{equation}
\Delta \nu = \nu_1(\hat{b}\cdot\hat{\sigma})(\hat{\epsilon}\cdot\hat{\epsilon}_s)+\nu_2(\hat{b}\cdot\hat{\epsilon}_s)(\hat{\epsilon}\cdot\hat{\sigma}),
\label{eqn:stark_interference_simple_vector}
\end{equation}
where $\hat{b}$ is the direction of the holding beam B-field, $\hat{\sigma}$ is the spin quantization axis, $\hat{\epsilon}$ is the holding beam polarization direction, and $\hat{\epsilon}_s$ is the direction of the applied static E-field.  Expressions for $\nu_1$ and $\nu_2$ are given in Appendix \ref{sec:stark_detail}.  We use the measured orientation of the applied field and their associated uncertainties to determine the $68.3\%$ confidence upper limits for the vector products in Eqn.\ \ref{eqn:stark_interference_simple_vector}.  We determine that  $(\hat{b}\cdot\hat{\sigma})(\hat{\epsilon}\cdot\hat{\epsilon}_s)\leq0.03$ and $(\hat{b}\cdot\hat{\epsilon}_s)(\hat{\epsilon}\cdot\hat{\sigma})\leq0.1$.

The false EDM signal is then obtained by calculating the phase shift this effect would cause in our EDM measurement using Eqn.\ \ref{eqn:phase_to_edm}.  This gives a 1-$\sigma$ uncertainty of $6\times10^{-26}$ $e$ cm for this systematic effect.  In the $\alpha$ scenario, this reduces to $2\times10^{-27}$ due to better field alignment. 

This effect is further suppressed in a retro-reflected holding beam geometry up to the power imbalance of the initial and retro-reflected beams.  In the $\beta$ scenario, we expect this systematic to appear at the $2\times10^{-28}$ $e$ cm level assuming a power imbalance of the holding beam at the $10\%$ level.  The size of this effect will decrease proportional to improvements in holding beam power balance.  To sufficiently correct for this systematic, we will need to directly evaluate Stark interference by varying the holding beam power as discussed in Section \ref{Holding_power_correlation}.  

It is possible to evaluate this systematic below the statistical uncertainty of the EDM measurement by first evaluating Stark interference in an intense traveling wave trap for both the co-magnetometer atom and for \Ra.  We estimate that Stark interference will cause a phase shift that corresponds to a false EDM signal at the $10^{-25}$ $e$ cm level for this geometry if we use the current maximum laser power of 30 W. This will provide a ratio measurement at the $0.1\%$ level of the effect of Stark interference in the co-magnetometer to that in \Ra.  Then we can measure the EDM-like phase shift of the co-magnetometer atom versus holding beam power in the retro-reflected geometry and use the precisely measured ratio to determine the effect of Stark interference on the EDM measurement with the precision of a co-magnetometer measurement, which is $3\times10^{-29}$ $e$ cm.

\subsection{\label{Geometric_phase}Geometric phase}

Spins moving in an inhomogeneous magnetic field experience a shift to their Larmor frequency due to the accumulation of geometric phases.  Part of this frequency shift can be proportional to the applied E-field and provide a false EDM signal.  The worst case scenario occurs when atoms take a peripheral orbit  in the trap and thus trace out a large area with their orbit.  In our experiment, we have a thermal cloud of atoms and must average over all types of orbits.  Using the treatment in Ref. \cite{Pendlebury04}, the false EDM signal from geometric phases is given by
\begin{equation}
d_\mathrm{false}=\frac{-F\hbar}{2B_{0z}^2 c^2}\left|v_{xy}\right|^2\frac{\partial B_{0z}}{\partial z}\frac{1}{1-\omega_r^2/\omega_0^2},
\label{eqn:geometric_phase}
\end{equation}
where $F=1/2$ is the total spin, $\left|v_{xy}\right|=\sqrt{2/3}v_\mathrm{D}$ is the RMS speed in the $\hat{x}-\hat{y}$ plane, $B_{0z}$ is the magnitude of the applied B-field, $\omega_0$ is the Larmor frequency, and $\omega_r$ is the trap frequency.  For the gradient in $B_{0z}$ we use $0.1\%$/cm which is the measured upper limit on the gradient of the B-field magnitude for displacement along $\hat{z}$.  The trap is asymmetric in the $\hat{x}-\hat{y}$ plane with trap frequencies $\omega_x=4$ Hz and $\omega_y=610$ Hz.  We calculate the uncertainty in this effect for $\omega_x$ because it is closest to $\omega_0$ and produces the largest effect.  The absolute value places an upper bound of $7\times 10^{-30}$ $e$ cm, which we use as an estimate of the 1-$\sigma$ uncertainty in this effect.  
In the $\beta$ scenario which uses a retro-reflected holding beam geometry, this reduces to $5\times 10^{-33}$ $e$ cm due to the changes in trap frequencies since cancellation of this effect improves as the trap frequencies and the precession frequency become more different.    

\subsection{\label{Systematics_Summary}Summary}

Table \ref{table:systematics} summarizes the current and projected 1-$\sigma$ uncertainties of all the systematic effects discussed above.  The total systematic uncertainty of $2\times10^{-25}$ $e$ cm is the quadrature sum of uncertainties for all individual effects.  The total systematic uncertainty is added in quadrature with the statistical uncertainty to calculate the $95\%$ confidence upper limit, although the systematic uncertainty is insignificant for this calculation.
The systematic uncertainty reported here represents an upper limit for the current measurement.  We have shown in the above discussion that for future EDM measurements with improved statistical uncertainty, we can limit the total systematic uncertainty to below the statistical uncertainty of the measurement.  
Improved field alignment in the $\alpha$ scenario and the addition of a co-magnetometer in the $\beta$ scenario allow ever more precise evaluation of systematic effects.    
It is important to stress that, even in the $\beta$ scenario, the listed uncertainties do not represent fundamental limitations.  Through more careful evaluation of the apparatus, co-magnetometry, and direct evaluation of certain systematic effects, we will be able to further reduce the uncertainty of these effects in an EDM measurement as the statistical sensitivity of the experiment improves.

\begin{table*}
\caption{\label{table:systematics} A list of systematic effects and the associated 1-$\sigma$ uncertainties for each effect in units of $e$ cm. No corrections are added in to the measured EDM value due to the small size of the effects and the dependence on conservative theoretical models.  Rather, we calculate the $68.3\%$ confidence upper limit for all effects and take this value as the 1-$\sigma$ systematic uncertainty.  Confidence limits are calculated from mean values and uncertainties of physical measurements that are used in the calculations of systematic effects.  The third and fourth columns list projected uncertainties for the $\alpha$ and $\beta$ scenarios, respectively.   Projected uncertainties that are labeled ``N/A'' do not need to be individually considered when using a co-magnetometer. We would observe these effects as an E-field correlated change in B-field and we would directly limit these effects with our limit on the B-field correlations systematic effect.}

\begin{ruledtabular}
\begin{tabular}{lccc}
Effect & Current Uncert. & $\alpha$ Scen. Uncert.  & $\beta$ Scen. Uncert.\\
\hline \\
E-squared effects & $1\times10^{-25}$ & $7\times10^{-29}$ &  $7\times10^{-31}$ \footnotemark[1]\\
B-field correlations & $1\times10^{-25}$ & $5\times10^{-27}$ & $3\times10^{-29}$ \footnotemark[1]\\
Holding ODT power Correlations & $6\times10^{-26}$ &  $9\times10^{-30}$ & $9\times10^{-32}$ \footnotemark[1]\\
Stark Interference & $6\times10^{-26}$ & $2\times10^{-27}$ & $3\times10^{-29}$ \footnote[1]{This uncertainty will improve with the statistical sensitivity of the experiment}\\
Blue laser power correlations & $7\times10^{-28}$ & $1\times10^{-31}$ & $1\times10^{-31}$ \\
Blue laser freq. correlations & $4\times10^{-28}$ & $8\times10^{-30}$ & $8\times10^{-30}$ \\
\textbf{E} x \textbf{v} effects & $4\times10^{-28}$ & $7\times10^{-30}$ & N/A \\
Leakage current & $3\times10^{-28}$ & $9\times10^{-29}$  & N/A\\
E-field Ramping & $9\times10^{-28}$ & $2\times10^{-29}$ & N/A\\
Geometric phase & $3\times10^{-31}$ & $7\times10^{-30}$ & $5\times10^{-33}$ \\
\hline
\\
Total & $2\times10^{-25}$ & $5\times10^{-27}$ & $4\times10^{-29}$ \footnotemark[1]\\
\end{tabular}
\end{ruledtabular}
\end{table*}

\section{\label{sec:Discussion}Discussion and Outlook}
This measurement of the \Ra\ EDM represents a significant improvement over our previous result \cite{Parker15} and demonstrates the progress that our experiment is poised to achieve.  
However, the achieved upper limit of $\left|d\left(^{225}\mathrm{Ra}\right)\right|\leq 1.4\times10^{-23}$ $e$ cm still does not improve current limits on \textit{CP} violation in the nuclear medium \cite{Chupp15} that are derived from measurements of the $^{199}$Hg EDM \cite{Griffith09,Graner16} and the neutron EDM \cite{Baker06}.  
However, there are several major experimental upgrades under development that can dramatically improve future measurements of the \Ra\ EDM.  

One limiting aspect of the current experiment is the small number of photons per atom that scatter out of the detection laser.  On average, it is possible to get 3 photons per atom before an atom decays to the dark state.  With so few photons scattered, it is unsurprising that the current measurement is limited by photon shot noise in the detection laser.  An increase in the number of photons scattered per atom would increase the signal size proportionally.  Since laser photon shot noise is the dominant noise source, increasing the signal size does not increase noise until we become limited by quantum projection noise (QPN), a fundamental source of noise resulting from projecting a quantum superposition state onto one of the basis states.  In this measurement, the laser light is either scattered out of the beam or not, corresponding to projecting the atom into either the bright or dark state, respectively.  

We can use the \Ss, $F=1/2$ to \Ps, $F=3/2$ (blue cycling) transition to detect atom number, which can scatter up to 1000 photons per atom on average before the atoms decay to a long-lived metastable state.  However, both the dark and bright nuclear spin states scatter light resonant on this transition so although this state is useful for detecting atom number, it cannot directly detect spin precession.  In future measurements, we plan to implement a nuclear-spin-dependent coherent transfer to the metastable $^3$D$_1$ state using stimulated Raman adiabatic passage (STIRAP) \cite{Bergmann98}.  This will allow us to detect spin precession using the blue cycling transition and improve the statistical sensitivity of an EDM measurement by over an order of magnitude.

With each atom contributing more signal to the measurement, the remaining fundamental limit to the statistical sensitivity of our experiment will then be QPN.  Sensitivity to an EDM is greatest if we measure for phase shifts in Larmor precession when the slope of population fraction versus precession time is greatest.  This corresponds to interrogating the atoms when they are in an equal superposition of the bright and dark states.  The QPN associated with projecting this state onto either the bright or dark state is proportional to the square root of the atom number while the signal is proportional to the atom number.  Thus, the statistical sensitivity of our measurement improves as the square root of the number of atoms we interrogate.

In future measurements, we can significantly gain in statistical sensitivity by increasing  the number of atoms we interrogate by many orders of magnitude.  First, we can address the disparity between the velocity of atoms exiting the oven and the capture velocity of the Zeeman slower by developing a longitudinal slower based on the blue cycling transition.  Although it is straightforward to design a variable magnetic field for atom deceleration with the blue cycling transition, keeping the three necessary repump lasers on resonance throughout the slower will add additional complexity to the slower operation because their Zeeman shifts are not matched to that of the blue cycling transition.  Effective atom slowing based on this transition is currently being developed using chirped frequency ramps rather than a variable magnetic field, a technique first demonstrated with sodium atoms \cite{Prodan84}.   

Furthermore, as new isotope production capabilities begin operation, the amount of \Ra\ that we load into the oven can be substantially increased.  
In the near future, efforts within the National Isotope Development Center to produce more $^{225}$Ac for the medical industry will also produce more \Ra.  Further in the future, efforts to harvest rare isotopes produced at the upcoming Facility for Rare Isotope Beams \cite{Pen14} have the potential to provide several orders of magnitude more \Ra\ than current production methods.

In addition to improvements in atom detection and increases in atom number, we are also fabricating new electrodes to achieve higher E-fields.  The current electrodes initially demonstrated a peak E-field of 100 kV/cm but now routinely only produce 65 kV/cm.  In contrast, other groups have demonstrated titanium electrodes that produce 800 kV/cm with no dark current using standard surface preparation techniques and a similar geometry to the current electrodes \cite{Furuta05}.  

In the QPN limited regime, a simple estimation for the statistical sensitivity of an EDM measurement is given by
\begin{equation}
\sigma_{EDM}=\frac{\hbar}{2E\sqrt{\tau N T}},
\label{eqn:stat_sens}
\end{equation}
where $\sigma_{EDM}$ is the standard error of the EDM measurement, $E$ is the magnitude of the applied E-field, $\tau$ is the precession time, $N$ is the number of atoms, and $T$ is the total measurement time.  An attractive route to achieving a statistical sensitivity of $1\times10^{-28}$ $e$ cm is to leverage new cooling and slowing techniques based on the blue transition and increases in the amount of available \Ra\  such that we increase the number of interrogated atoms from 500 to $5\times10^6$ atoms.  With improved electrodes, we would then need to achieve a field of 150 kV/cm to reach $1\times10^{-28}$ $e$ cm with an integration time of 60 days.

The combination of improving atom detection, increasing atom number, and increasing E-field strength form a solid framework for future \Ra\ EDM measurements that are many orders of magnitude more precise than the current results.  Furthermore, there are no known systematic effects that would prohibit a measurement at this level.  Taking into account the unique sensitivity \Ra\ has to \textit{CP} violation in the nuclear medium, a measurement at the $10^{-28}$ $e$ cm level would probe \textit{CP} violation with unprecedented precision.  

This work is supported by U.S. Department of Energy (DOE), Office of Science, Office of Nuclear Physics, under contracts No. DE-AC02-06CH11357 and No. DE-FG02-99ER41101.  \Ra\ used in this research was supplied by DOE, Office of Science, Isotope Program in the Office of Nuclear Physics. M. B. acknowledges support from Argonne Director’s postdoctoral fellowships.

%

\appendix

\section{\label{appendix:averaging}Data Averaging}

We use the term ``cycle'' to refer to a single implementation of the experiment consisting of MOT loading, atom transfer, and atom detection.
For each cycle we produce five shadow magnitudes from the five atom detection images, however, image 1 is currently not used for analysis.  
We operate the experiment over multiple cycles without changing $\delta$.   
We accumulate shadow magnitudes taken with the same $\delta$ and E-field condition from multiple cycles (and their associated normalization images) into ``bins'' and calculate the mean and standard error of the image data in each bin.  
The sign of the E-field is alternated each cycle, so bins contain data from every other cycle for images 2 and 3 and data from each cycle for images 4 and 5, because no E-field is applied during the precession time before image 4.  
The number of shadow magnitudes in each bin is known as the ``bin size'' and we analyze the data using bin sizes from 3 to 64, which is typically the maximum number of cycles that occur before we change $\delta$.  
A ``set'' of data refers to the data from multiple cycles that are taken during continuous operation of the experiment while no changes to $\delta$ are made. 
We never collect data from different sets into the same bin even if both sets are taken with the same values of $\delta$.  

 This binning procedure is useful to optimize the statistical sensitivity of our measurement because we have known sources of non-Gaussian noise that occur at different time scales.  On one hand, making bins larger reduces  sensitivity to power fluctuations in the blue laser.  The power is manually adjusted between sets to scatter 2.1 photons per bright state atom, however, within a set the power is observed to fluctuate within $\pm10\%$ of its initial value.  Since the time scale for these drifts is similar to the cycle time of 100 s, collecting more cycles together in one bin reduces sensitivity to these fluctuations.  On the other hand,  drifts in atom number due to oven temperature changes and oven depleting occur at a much longer time scale of several hours to days.  By setting the bin size small enough such that the atom signal is normalized before the change in atom number becomes significant, we can reduce sensitivity to non-random fluctuations in atom number.  We find that a bin size of 32 is optimal based on the reduced chi-square of the fit to the model (Eqn.\ \ref{eqn:fit_equations}).  This corresponds to collecting all data with E-field applied into a single bin and separating the data with no E-field applied into two bins for the most common data set size of 64 experimental cycles.  This shows that separation into data sets is already sufficient to eliminate sensitivity to non-random fluctuations in atom number.

\section{\label{sec:stark_detail}Stark Interference Details}

We perform a similar calculation of this effect to one for $^{199}$Hg \cite{Romalis99}, where the energy shift of the $m_F$ ground state Zeeman sublevel is given by 
\begin{widetext}
\begin{equation}
\begin{array}{lcl}
\Delta \mathbb{E}(m_F) & = & \frac{\mu_B E_0^2 E_s}{4c\hbar^2}\sum_{S,F',F'',m_{F'},m{F''}}\mel{^{2S+1}\mathrm{P}_1,F',m_{F'}}{\hat{\mathbf{\epsilon}}\cdot \mathbf{r}}{{}^1\mathrm{S}_0,F=1/2,m_F}\times \\
\\
& &\mel{^{2S+1}\mathrm{P}_1,F'',m_{F''}}{\hat{\mathbf{b}}^*\cdot \left(\mathbf{L}+2\mathbf{S}\right)}{{}^{2S+1}\mathrm{P}_1,F',m_{F'}}\times \\
\\
& &\mel{{}^1\mathrm{S}_0,F=1/2,m_F}{\hat{\mathbf{\epsilon}}_s\cdot \mathbf{r}}{^{2S+1}\mathrm{P}_1,F'',m_{F''}}\frac{1}{\left(\omega' - \omega_L\right)\left(\omega'' - \omega_L\right)}+\mathrm{perm.}+\mathrm{c.r.}
\label{eqn:stark_interference_energy}
\end{array}
\end{equation}
\end{widetext}
Here, $\mu_B$ is the Bohr magneton, $E_0$ is the holding beam electric field amplitude, $\hat{\epsilon}$ is the polarization vector of the holding beam, $E_s$ is the magnitude of the applied static E-field, $\hat{\epsilon}_s$ is the direction of the applied E-field, $\hat{b}$ is the direction of the holding beam magnetic field, and $\omega'('')$ is the frequency of the $\ket{^1S_0}$ to $\ket{^{2S+1}P_1, F'(''), m_{F'}('')}$ transition.  In principle, ``$\mathrm{perm.}$'' represents all six permutations of $\hat{\mathbf{\epsilon}}\cdot \mathbf{r}$, $\hat{b}\cdot \left(\mathbf{L}+2\mathbf{S}\right)$, and $\hat{\epsilon}_s\cdot \mathbf{r}$, however, magnetic dipole amplitudes within the ground state manifold are suppressed by the ratio of $\mu_B/\mu_N$, where $\mu_N$ is the nuclear magneton, so only the two permutations that have magnetic dipole amplitudes between excited state hyperfine levels are significant.  We also need to include counter-rotating terms, designated by ``$\mathrm{c.r.}$'' in Eqn.\ \ref{eqn:stark_interference_energy}. The sum over excited states is limited to the \Ps\ and \Pt{1}\ states since the electric dipole amplitudes between these states and \Ss\ dominate those of other states by many orders of magnitude.  

To simplify Eqn.\ \ref{eqn:stark_interference_energy}, we use a standard technique to reduce dipole matrix elements by expressing them in terms of rank 1 spherical tensor operators $T_q^{1}$ and employing the Wigner-Eckart theorem to factor out all angular dependence. The energy shift due to a specific combination of tensor operators $T_{q_\epsilon}^1$, $T_{q_s}^1$, and $T_{q_b}^1$ is then given by,
\begin{widetext}
\begin{equation}
\begin{array}{c}
\Delta E(m_F,q_\epsilon,q_s,q_b) = \frac{\mu_B E_0^2 E_s}{4c\hbar^2}\sum_{S,F',F'',m_{F'},m_{F''}}\left[\frac{(-1)^{q_b}G(m_F,F',m_{F'},F'',m_{F''},q_s,-q_b,q_\epsilon)}{(\omega'-\omega_L)(\omega''-\omega_L)} \right. \\
\\
\left. + \frac{(-1)^{q_\epsilon}G(\text{``\quad''},q_s,q_b,-q_\epsilon)}{(\omega'+\omega_L)(\omega''+\omega_L)} + \frac{(-1)^{q_\epsilon}G(\text{``\quad''},-q_\epsilon,q_b,q_s)}{(\omega'-\omega_L)(\omega''-\omega_L)} + \frac{(-1)^{q_b}G(\text{``\quad''},q_\epsilon,-q_b,q_s)}{(\omega'+\omega_L)(\omega''+\omega_L)} \right],
\label{eqn:stark_interference_factored_energy}
\end{array}
\end{equation}
where the function G is defined to be
\begin{equation}
\begin{array}{c}
G(m_F,F',m_{F'},F'',m_{F''},q_1,q_2,q_3) \equiv (2F'+1)(2F''+1)(-1)^{1+3F''+7I+J''+2J'-m_F-m_{F'}-m_{F''}+3F'} \\
\\
\times
\left\{\begin{array}{ccc}
J'' & 1 & J' \\
F' & I & F''  
\end{array}\right\}

\left(\begin{array}{ccc}
I & 1 & F'' \\
-m_F & q_1 & m_{F''}  
\end{array}\right)

\left(\begin{array}{ccc}
F'' & 1 & F' \\
-m_{F''} & q_2 & m_{F'}  
\end{array}\right)

\left(\begin{array}{ccc}
F' & 1 & I \\
-m_{F'} & q_3 & m_F  
\end{array}\right)

\left|\left\langle^1\text{S}_0\left|\left|^{}d_{}\right|\right|{}^{2S+1}\text{P}_1\right\rangle\right|^2\times

\left\{\begin{array}{cc}
\sqrt{2/3} & \text{if}\; S=0 \\
3/\sqrt{6} & \text{if}\; S=1  
\end{array}\right. .

\label{eqn:G_function}
\end{array}
\end{equation}
\end{widetext}

Here, curly brackets denote a Wigner 6-j symbol, parentheses denote a Wigner 3-j symbol, and the double vertical lines represent a reduced matrix element.  Values of the reduced matrix elements used in this calculation are taken from \cite{Dzuba06}.

Finally we group the energy shifts, expressed in the form of Eqn.\ \ref{eqn:stark_interference_factored_energy}, into a single shift if they have the same vector dependence.  Since only two nonzero vector components exist, we can express the total frequency shift between Zeeman sub-levels due to stark interference in terms of the two frequencies, $\nu_1$ and $\nu_2$, from Eqn.\ \ref{eqn:stark_interference_simple_vector}. These frequencies are given by

\begin{widetext}
\begin{equation}
\begin{array}{lcl}
\nu_1 &=& -\frac{1}{2h}\left(\left[\Delta E(1/2,-1,0,-1)+\Delta E(1/2,1,0,1)-\Delta E(1/2,-1,0,1)-\Delta E(1/2,1,0,-1)\right]-\right.\\
\\
& &\left.\left[\Delta E(-1/2,-1,0,-1)+\Delta E(-1/2,1,0,1)-\Delta E(-1/2,-1,0,1)-\Delta E(-1/2,1,0,-1)\right]\right)
\\
\\
&& \mathrm{and} \\
\\
\nu_2 &=&-\frac{1}{2h}\left(\left[\Delta E(1/2,0,-1,-1)+\Delta E(1/2,0,1,1)-\Delta E(1/2,0,-1,1)-\Delta E(1/2,0,1,-1)\right]\right.\\
\\
& &\left.\left[\Delta E(-1/2,0,-1,-1)+\Delta E(-1/2,0,1,1)-\Delta E(-1/2,0,-1,1)-\Delta E(-1/2,0,1,-1)\right]\right).
\label{eqn:nu1_and_nu2}
\end{array}
\end{equation}
\end{widetext}

\end{document}